\documentclass[english,
]{aa}
\usepackage{graphicx}
\usepackage{natbib}
\usepackage{txfonts}

\bibpunct{(}{)}{;}{a}{}{,} 

\begin{document}

\title{The dawn of a new era for dustless HdC stars with GAIA eDR3
\thanks{The spectra presented in Figs.~\ref{fig_Spectra_newHdC_cool},~\ref{fig_Spectra_newHdC_mild}, ~\ref{fig_Spectra_newHdC_warm} and ~\ref{fig_Spectra_eHe} are only available at the CDS via anonymous ftp to cdsarc.u-strasbg.fr (130.79.128.5) or via http://cdsarc.u-strasbg.fr/viz-bin/cat/J/A+A/vol/page}}

\author{
P.~Tisserand\inst{1},
C.~L.~Crawford\inst{2},
G.~C.~Clayton\inst{2},
A.~J.~Ruiter\inst{3},
V.~Karambelkar\inst{5}, 
M.~S.~Bessell\inst{4},
I.~R.~Seitenzahl\inst{3},
M.~M.~Kasliwal\inst{5},
J.~Soon\inst{4},
T.~Travouillon\inst{4},
}


\institute{
Sorbonne Universit\'es, UPMC Univ Paris 6 et CNRS, UMR 7095, Institut d'Astrophysique de Paris, IAP, F-75014 Paris, France \and
Department of Physics \& Astronomy, Louisiana State University, Baton Rouge, LA 70803, USA \and
ARC Future Fellow, School of Physical, Environmental and Mathematical Sciences, University of New South Wales,\\
Australian Defence Force Academy, Canberra, ACT 2600, Australia\and
Research School of Astronomy and Astrophysics, Australian National University, Cotter Rd, Weston Creek ACT 2611, Australia \and
Cahill Center for Astrophysics, California Institute of Technology, Pasadena, CA 91125, USA
}

\offprints{Patrick Tisserand; \email{tisserand@iap.fr}}

\date{}


\abstract {Decades after their discovery, only four hydrogen-deficient carbon (HdC) stars were known to have no circumstellar dust shell. This is in complete contrast to the $\sim$130 known Galactic HdC stars that are notorious for being heavy dust producers, that is the R Coronae Borealis (RCB) stars. Together, they form a rare class of supergiant stars that are thought to originate from the merger of CO/He white dwarf (WD) binary systems, otherwise known as the double-degenerate scenario.}
{We searched for new dustless HdC (dLHdC) stars to understand their Galactic distribution, to estimate their total number in the Milky Way, and to study their evolutionary link with RCB stars and extreme helium (EHe) stars, the final phase of HdC stars.}
{We primarily used the 2MASS and GAIA eDR3 all-sky catalogues to select candidates that were then followed-up spectroscopically. We studied the distribution of known and newly discovered stars in the Hertzsprung-Russell diagram.}
{We discovered 27 new dLHdC stars, one new RCB star, and two new EHe stars. Surprisingly, 20 of the new dLHdC stars share a characteristic of the known dLHdC star HD 148839, having lower atmospheric hydrogen deficiencies. The uncovered population of dLHdC stars exhibits a bulge-like distribution, like the RCB stars, but show multiple differences from RCB stars that indicate that they are a different population of HdC stars. This population follows its own evolutionary sequence with a fainter luminosity and also a narrow range of effective temperatures, between 5000 and 8000 K. Not all the new dLHdC stars belong to this new population, as we found an indication of a current low dust production activity around 4 of them: the warm F75, F152, and C526, and the cold A166. They might be typical RCB stars passing through a transition time, entering or leaving the RCB phase.}
{For the first time, we have evidence of a wide range of absolute magnitudes in the overall population of HdC stars, spanning more than 3 mag. In the favoured formation framework, this is explained by a wide range in the initial total WD binary mass, which leads to a series of evolutionary sequences with distinct maximum brightness and initial temperature. The cold Galactic RCB stars are also noticeably fainter than the Magellanic RCB stars, possibly due to a difference in metallicity between the original population of stars, resulting in a different WD mass ratio. The unveiled population of dLHdC stars indicates that the ability to create dust might be linked to the initial total mass. In our Galaxy, there could be as many dLHdC stars as RCB stars.}


\authorrunning{Tisserand, P. }
\titlerunning{Dawn of a new era for dustless HdC stars with GAIA eDR3}

\maketitle

\section{Introduction \label{sec_intro}}

Hydrogen-deficient carbon (HdC) stars form the class of rare supergiant stars whose best-known members are the R Coronae Borealis (RCB) stars \citep{1996PASP..108..225C,2012JAVSO..40..539C,2020A&A...635A..14T}. RCB stars are notorious for their surprising photometric variabilities as they undergo unpredictable large declines in brightness over a few weeks (up to 9 mag in the visible) before recovering their original brightness in a few months (see \citealp{2008A&A...481..673T, 2009A&A...501..985T, 2013A&A...551A..77T} for some examples of light curves). These brightness variations are due to clouds of carbon dust ejected by the stars along the line of sight that obscure the photosphere. Because this occurs continuously, amorphous carbon dust grains \citep{2011ApJ...739...37G} build up around the stars, and circumstellar dust shells form \citep{2012A&A...539A..51T,Montiel_2015}. Five members of the HdC stars formed a small subclass for decades as no variability was ever reported, even though they have the same peculiar spectroscopic characteristic as RCB stars \citep{1967MNRAS.137..119W}. \citet{1997MNRAS.285..317F} called them the non-RCB HdC stars, but they were called simply HdC stars in many articles. They are \object{HD 137613}, \object {HD 148839}, \object{HD 173409}, \object{HD 182040} and \object{HD 175893}. For the first four, \citet{2012A&A...539A..51T} reported that no circumstellar dust shell has been detected with the WISE mid-infrared space telescope \citep{2010AJ....140.1868W}, supporting the fact that no carbon dust is currently being ejected from their atmospheres. From now on, we call them the dustless HdC (dLHdC) stars to avoid any confusion with the HdC stars, their parent class. The last star, HD 175893, was found to possess a warm circumstellar dust shell, similar to that of RCB stars, indicating an on-going phase of dust production. This dust production did not coincide with a visual photometric decline. The reason is not known; it might due to a particular location of the dust production site that prevents the dust from reaching the line of sight.    
   
Historically, HD 137613 and HD 173409 were reported nearly 130 years ago to be stars with peculiar spectra by \citet{1892AstAp..11..765F} and \citet{1896ApJ.....4..142P}, respectively, while HD 182040 was mentioned by \citet{1912ApJ....35..125P} to stand in a class by itself, while reporting Mrs Fleming observation that none of the lines in the spectrum of HD 182040 are those due to hydrogen. Then \citet{1953ApJ...117...25B} mentioned that the four following stars, HD 137613, HD 173409, HD 175893, and HD 182040 have similar spectra as the known variable RCB stars, but did not show any brightness variations. Bidelman described spectra showing characteristic strong C$_2$ band heads and some C I lines indicating an atmosphere rich in carbon, made mostly of isotope 12 as no evidence of the carbon isotope 13 was found. They are also hydrogen poor as no CH bands and no hydrogen absorption lines were detected. Ten years later, \citet{1963MNRAS.126...61W} reported a fifth star, HD 148839, which belongs to the group of hydrogen-poor carbon stars, but with a hydrogen deficiency that is not as marked as that of the first four. Since then, only one dLHdC star has been reported. This is \object{HE 1015-2050}, a faint (G$\sim$16 mag) HdC star located in the Galactic halo \citep{2010ApJ...723L.238G}. However, although no IR excess has been observed from the 2MASS or WISE catalogues\footnote{2MASS: (J-H)$_{0}\sim$0.20 mag and (H-K)$_{0}\sim$0.27 mag, WISE: [3.4]-[4.6]$\sim$0 mag, only magnitude limit values were reported for the reddest WISE bands, [12] and [22], in both WISE All-Sky and ALLWISE catalogues}, indicating that no warm circumstellar dust shell has built up in large quantity around this star, \citet{2013ApJ...763L..37G} observed similar spectroscopic emission features as are usually seen in RCB stars that undergo photometric declines. They also inferred the presence of circumstellar material from their polarimetric observations. These are really interesting observations as HE 1015-2050 could be useful for understanding the evolutionary link between the two types of HdC stars. Nevertheless, as no photometric variability has been observed in the past 16 years from the Catalina \citep{2012IAUS..285..306D} and ASAS-SN \citep{2014ApJ...788...48S,2017PASP..129j4502K} monitoring surveys, we continue to classify HE 1015-2050 as a dLHdC star. 

The distances and G-band brightnesses reported by the GAIA survey \citep{2016A&A...595A...1G} for the first four dLHdC stars range between 1 and 2 kpc, and between 6th and 9th magnitude, respectively. There was clearly room for progress in increasing their numbers, especially as an important successful effort has been made in the past 20 years to search for new RCB stars up to the distance of the Galactic bulge \citep{2005AJ....130.2293Z, 2008A&A...481..673T, 2011A&A...529A.118T, 2013A&A...551A..77T, 2019MNRAS.483.4470S, 2020A&A...635A..14T, 2021ApJ...910..132K} and in the Magellanic Clouds \citep{1996ApJ...470..583A,2001ApJ...554..298A,2004A&A...424..245T, 2009A&A...501..985T}. The search for new dLHdC stars is motivated by the wish to understand their Galactic distribution and to test the first estimate of their total Galactic numbers made by \citet{2020A&A...635A..14T}. Based on only the brightest four known dLHdC stars, it was estimated that up to one dLHdC star might exist for every six RCB stars. Clearly, this first estimate has a large uncertainty because the initial sample was small. It is important to evaluate this ratio accurately as it is used to estimate the total number of Galactic HdC stars, giving a strong constraint on their formation rate. As HdC stars are thought to result from the merger of one carbon-oxygen (CO) plus one helium (He) white dwarf (WD) \citep{1984ApJ...277..355W,2011MNRAS.414.3599J,2012JAVSO..40..539C}, these constraints are used to test the double-degenerate scenario when compared to the birthrate of these mergers obtained by theoretical population synthesis, that is, between $\sim10^{-3}$ and $\sim5\times10^{-3}$ per year \citep{2001A&A...365..491N,2009ApJ...699.2026R,2015ApJ...809..184K}. 

Finally, we refer to the complementary analyses made by \citet{crawford_2022} and \citet{Karambelkar_2022}, which were carried out on the optical and near-IR spectra of some of these new dLHdC stars, respectively. The first paper studied the astonishing strong strontium and barium absorption lines we found in A249, C539, and A166 and used stellar evolution models to estimate the type of neutron exposure that occurs in a typical HdC star to explain the observed enhanced s-process material on their surfaces. The second analysis measured the oxygen isotopic ratio $^{16}$O/$^{18}$O of all new cold dLHdC stars for which the CO band heads were detectable in the K band and compared them to values obtained for RCB stars. They confirmed the lower $^{16}$O/$^{18}$O ratio observed in dLHdC stars as indicated by \citet{2007ApJ...662.1220C} and \citet{2009ApJ...696.1733G}, who suggested that dLHdC stars have $^{16}$O/$^{18}$O$<$1 and most RCB stars have $^{16}$O/$^{18}$O$>$1.

We searched for new dLHdC stars using the datasets of two all-sky surveys:  2MASS \citep{2006AJ....131.1163S}, and GAIA eDR3 \citep{2021A&A...649A...1G}. We limited ourselves to stars whose lines of sight are affected by an integrated interstellar reddening E(B-V) $<$ 1 mag \citep{2011ApJ...737..103S}. The selection of candidates based on their brightness and colour is detailed in section~\ref{sec_Analysis}, then in section~\ref{sec_discoveries} we describe the spectroscopic follow-up of these candidates and the subsequent discovery of new dLHdC, RCB, and EHe stars. In section~\ref{sec_result} we discuss the properties of this population of dLHdC stars and compare them to the known Galactic RCB stars in spatial distribution, brightness, effective temperature, photometric variability, dust production rate, and spectral energy distribution. We also estimate their total number in the Milky Way, before summarising our results in section~\ref{sec_summary}.

\section{Analysis \label{sec_Analysis}} 

\begin{figure*}
\centering
\includegraphics[width=7in,origin=c]{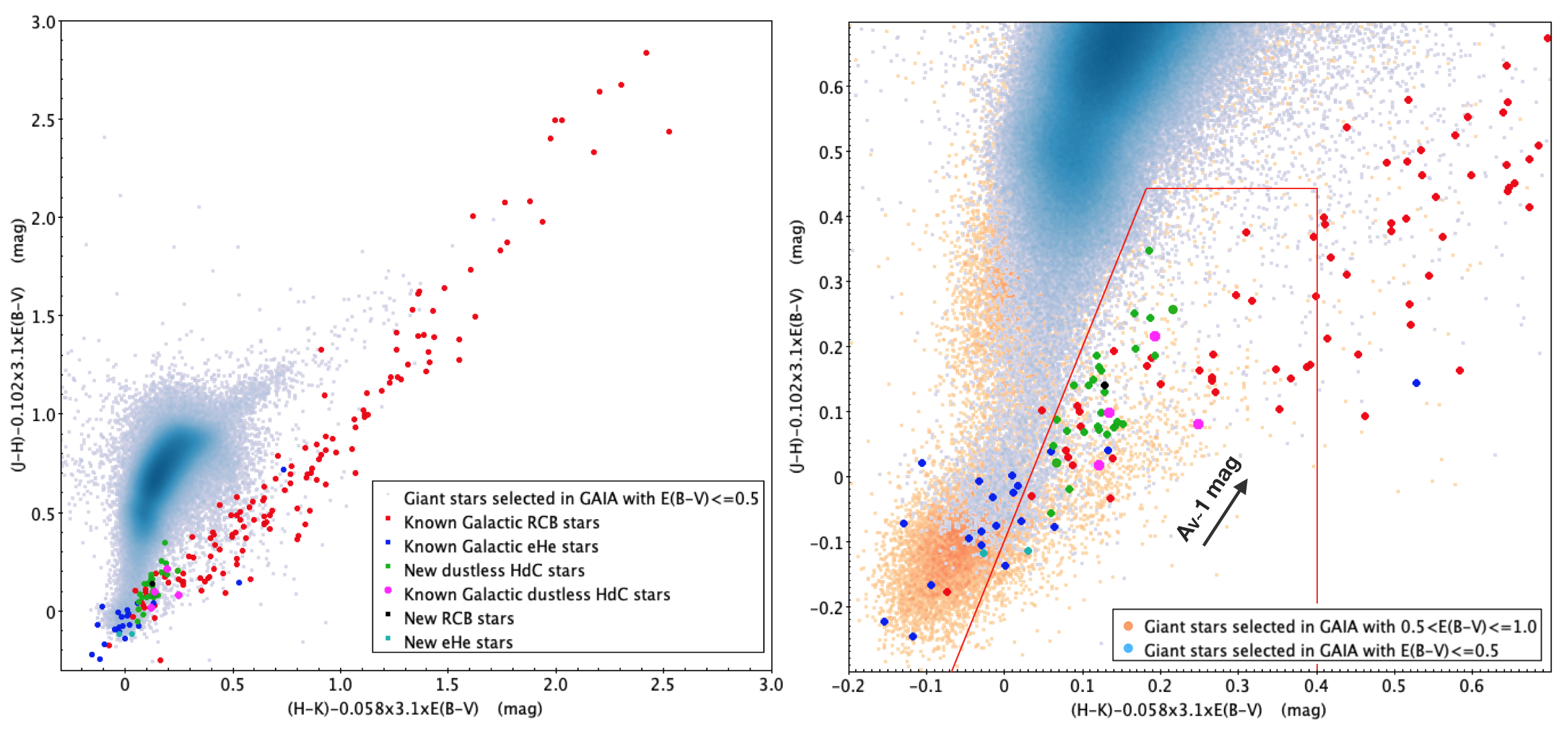}
\caption{Colour-colour (J-H)$_0$ vs (H-K)$_0$ diagram with 2MASS magnitudes corrected for extinction, showing the positions of all known and new Galactic HdC and EHe stars, represented with colour points, in comparison to giant stars (M$_{G}\leq$0 mag, BP-RP$>$0.5 mag, G$<$15 mag) selected in GAIA eDR3 (groups A and B). Left: Distribution of all giant stars with an interstellar reddening E(B-V) lower than 0.5 mag. Right: Zoom on the lower left corner. The distribution of giant stars impacted by interstellar reddening between 0.5 $<$ E(B-V) $\leq$1.0 mag is indicated in red, and the distribution of the ones of lower reddening is shown in blue. The selection area is delimited with red lines. The interstellar dust reddening vector is represented for an extinction A$_{V}$ of 1 mag.}
\label{fig_Select-2MASS}
\end{figure*}

\subsection{Initial star lists \label{sec_starlist}}

With the release of the GAIA eDR3 all-sky catalogue, a large-scale search of rare objects such as dLHdC stars is now possible as supergiant stars can be selected up to the distance of the Galactic bulge thanks to the reported precise parallax measurements. Then, as an initial selection, we downloaded the data for all giant stars with a GAIA M$_G$ $<$0 mag, BP-RP$>$0.5 mag, an apparent G magnitude brighter than 15th mag and an estimated parallax signal-to-noise ratio (S/N) higher than 3. We used here, for simplification, the inverse of the GAIA eDR3 parallax as a first distance estimate. The catalogue was simply cleaned of possible incorrect parallax measurements using the two following constraints on GAIA eDR3 parameters, RUWE$\leq$1.4 or ipd\_gof\_harmonic\_amplitude$\leq$0.1, as recommended by \citet{2021A&A...649A...2L}. HE 1015-2050, which is fainter than 15th$^{}$ magnitude and for which no significant parallax measurement is reported by GAIA eDR3, is the only known dLHdC star that did not fully pass these initial criteria.

Next, we cross-matched this sub-sample of GAIA stars with the ALLWISE catalogue \citep{vizier:II/328} using the GAIA archive service (https://gea.esac.esa.int/archive/). This catalogue contains the WISE mid-infrared magnitudes and also the 2MASS J, H, and K near-IR magnitudes as an internal cross-matching was already performed. However, not all GAIA stars were matched to an ALLWISE counterpart, so that we also cross-matched the remaining GAIA stars to the 2MASS catalogue with a 1$\arcsec$~radius association criterion
using the CDS cross-matching tools (http://cdsxmatch.u-strasbg.fr/). This was the case for the known dLHdC star, HD 182040. We also calculated an E(B-V) interstellar reddening magnitude for
all selected GAIA stars following \citet{2011ApJ...737..103S}. The GAIA BP-RP colour and G brightness were corrected using these simple formulas: $\bigtriangleup$(BP-RP)$\sim2\times$E(B-V) for the reddening correction, and A$_{G}\sim$3.1$\times$E(B-V) for the visual extinction \citep{2018A&A...616A...8A,2018MNRAS.481.3442M}. For simplicity, we kept our search for new dLHdC stars in the sky area impacted by an interstellar reddening E(B-V) lower than 1 magnitude.

For the definition of the subsequent selection criteria on absolute brightness and colour, we categorised our initial samples of GAIA stars detailed above into four groups. Groups A and C contain the giant stars that have a successful match with the WISE ALLWISE catalogue, but were reported with a parallax S/N higher than 5, and between 3 and 5, respectively. Groups B and D were defined with the same split on parallax, but only with the stars for which no association with the WISE ALLWISE catalogue was found and a cross-matching with the 2MASS catalogue was then necessary.

Two more groups, groups F and G, were defined to open up our search to GAIA stars reported with no significant distance estimate (parallax S/N lower than 3) towards sky areas impacted by an interstellar reddening E(B-V)<0.5 mag. This initial set of stars also followed the criteria defined above on the GAIA G apparent brightness (G<15 mag) and colour (BP-RP$>$0.5 mag).  Similar to the first four groups, group F corresponds to stars with a WISE ALLWISE catalogue counterpart, while stars from group G only have a 2MASS association.

Overall, we started our analysis with an initial set of about 3.5 million GAIA stars, $\sim$85\% of which were located on lines of sight with median interstellar dust reddening E(B-V) lower than 0.5 mag. Details on the distribution of these stars among the different groups is given in Table~\ref{tab.Selection}.

\begin{figure*}
\centering
\includegraphics[width=7in,origin=c]{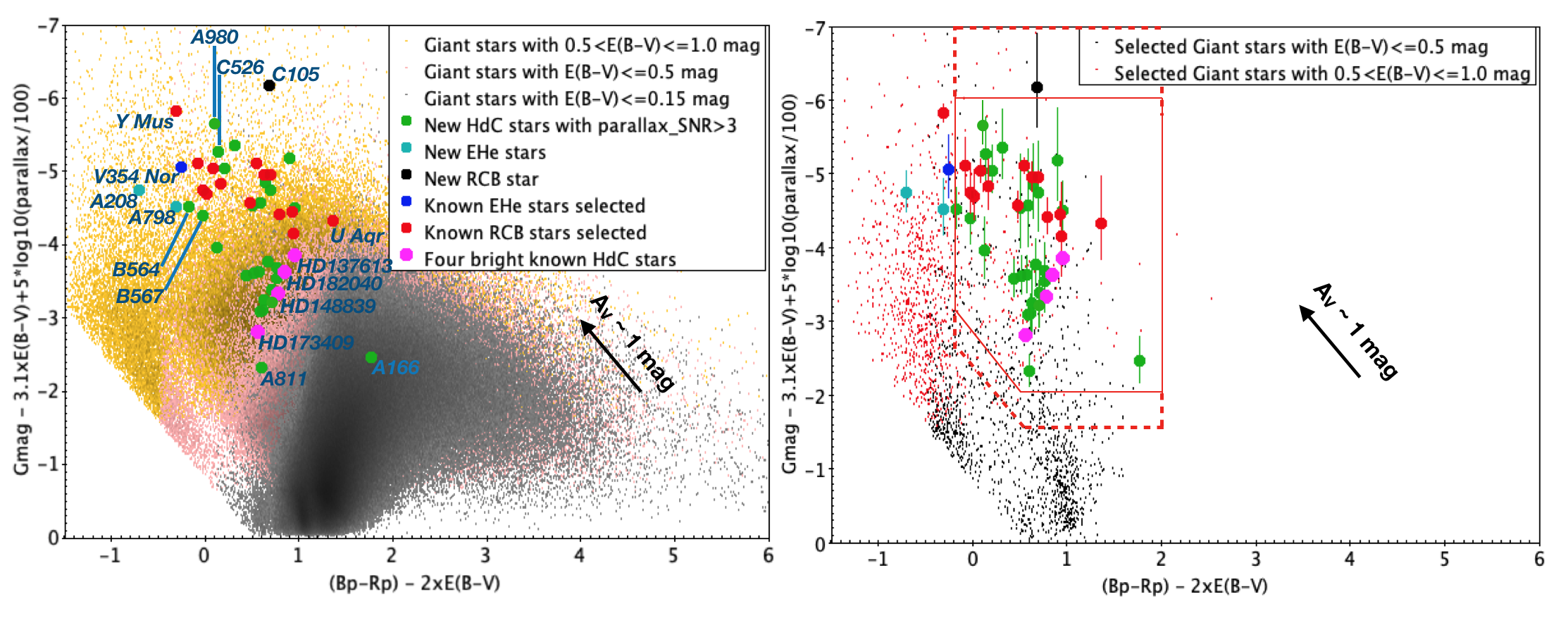}
\caption{HR diagram M$_{G}$ vs (BP-RP) with GAIA magnitudes corrected for interstellar reddening. Left: Distribution of all giant stars whose measured parallax S/N is higher than 5 (groups A and B), but that are impacted by a low level of interstellar reddening (E(B-V)$\leq$ 0.15 mag, grey dots), a medium level (E(B-V)$\leq$0.5 mag, pink dots), or an even higher level (0.5$<$E(B-V)$\leq$1.0 mag, orange dots). The positions of the known HdC and EHe stars that have passed the first selection cuts are indicated, as well as the newly discovered ones. Right: Same as the left, except that the represented giant stars are those that have passed the first selection cuts. The black dots correspond to the objects impacted by a medium value of interstellar reddening, E(B-V)$\leq$0.5 mag, while the red dots show those impacted by higher values up to 1.0 mag. The error bars on the y-axis due to the parallax measurement accuracy are shown only in the right diagram for clarity. The selection areas are marked with red lines: solid lines for groups A and B, and dashed lines for groups C and D, whose parallax is measured with an S/N lower than 5. The interstellar extinction of A$_{V}\sim$1 mag is represented with a black arrow.}
\label{fig_Select-GAIA}
\end{figure*}

\subsection{Selection criteria \label{sec_criteria}}

We primarily used the GAIA eDR3 and the 2MASS all-sky catalogues to select targets that we followed-up spectroscopically. The mid-IR WISE magnitudes were used to verify that no large fraction of selected stars belonging to groups A, C, and F possessed a clear circumstellar dust shell. We found that the  $[4]-[12]$ and $[12]-[24]$ colour indices of most stars that passed our first two primary selection cuts, detailed below, are distributed around zero. Therefore we did not apply strict criteria on their WISE colours to retain potential discoveries of new RCB stars, which usually show an excess because they are surrounded by warm dust. After the initial selection of giant stars from the GAIA eDR3 catalogue, we applied the three selection criteria described below. The overall selection process is summarised in Table~\ref{tab.Selection} for groups A to D.

First, we selected a first sample of stars whose position in the (J-H)$_0$ vs (H-K)$_0$ colour-colour diagram, corrected for interstellar reddening, was similar to the positions of the four brightest known dLHdC stars. The selected area corresponds to stars with no high IR excess, unlike most RCB stars. Fortuitously, the four known dLHdC stars are also separated from the main locus of classical giant stars with a redder H-K colour index. We therefore defined some specific limits that are illustrated in Figure~\ref{fig_Select-2MASS}. This selection has a high impact as it rejected about 99.7\% of the initial sample of giant stars. In practice, we selected stars using the following criteria with $\alpha$ = 0.1:

\begin{eqnarray}
&& if\hspace{1 mm}0.18<(H-K)_0< 0.4 :  (J-H)_0<0.44  \nonumber  \\
&& if\hspace{1 mm}(H-K)_0\leqslant 0.18 :  (J-H)_0<3\times(H-K)_0-\alpha
\label{eq.cut1}
.\end{eqnarray}  

Second, we used the GAIA eDR3 M$_{G}$ versus (BP-RP)$_0$ colour-magnitude diagram to select stars whose brightness and colour were similar to those of the four known dLHdC stars. The selection area is illustrated in Figure~\ref{fig_Select-GAIA}, right side. As we chose, for simplicity, to use the inverse of the GAIA eDR3 parallax as a distance estimate, we relaxed our cuts on the brightness scale. A more accurate geometric distance estimate inferred by \citet{2021AJ....161..147B} is used is Section~\ref{sec_cmd} to study the population of dLHdC stars in details. Overall, the limits were chosen pragmatically, with the goal to deliberately keep the range of colour and brightness as wide as possible while maximising the rejection of two groups of stars: the bright blue clump of stars, and the group of fainter giant stars whose colour (BP-RP$\sim$1.0 mag) is similar to that of the stars of the red clump, which are one magnitude fainter \citep[see][Fig.10]{2018A&A...616A..10G}. The selection limits were chosen to be even wider on the brightness scale for groups C and D because their respective distances are less well known. About 12\% of the stars that passed cut 1 were selected. It is interesting to note that the four known dLHdC stars that we used as reference in our analysis lie at a position of the HR diagram that is underpopulated when we compare them to some giant stars that are impacted by very low interstellar reddening, that is, E(B-V)$\leq$0.15 mag (see Figure~\ref{fig_Select-GAIA}, left side). The four known dLHdC stars are brighter and bluer than the main locus of stars that is mainly formed by the asymptotic giant branch (AGB) stars. In sky areas of higher interstellar reddening, the imprecision of the knowledge of the interstellar extinction has the direct consequence of dispersing the distribution of stars in the HR diagram. We can therefore expect that most of the selected stars will be AGB stars, but some hot blue stars whose extinction correction was underestimated will also be selected.

Finally, for the third and last selection, we queried the CDS/SIMBAD database \citep{2000A&AS..143....9W} for information about all stars that passed the selection cuts described above. We then rejected all targets that already possessed a well-defined spectroscopic type that is not carbon rich, and all targets associated with a globular cluster. We also removed all known dLHdC stars and RCB stars that have passed all the applied criteria; there were 18 of them, as listed in Table~\ref{tab.Selection}. About 38\% of the stars selected in cut 2 were removed at this stage from the final target list. We selected 631 stars from groups A to D for spectroscopic follow-up, $\sim$44\% of which were located in a sky area of low interstellar reddening, that is, E(B-V)$\leqslant$0.5 mag. As we started with an initial dataset in which these stars represented $\sim$85\% of the total, our selection process was less efficient with stars impacted by higher extinction.

For the nearly 400 thousand stars belonging to groups F and G, whose measured parallax S/N is lower than 3, we applied a stricter selection on their 2MASS near-IR colours with $\alpha$=0.2 in Eq.~\ref{eq.cut1} because of a higher dispersion. We also requested a tighter constraint on their GAIA (BP-RP)$_0$ colour, ranging between 0.4 and 1.2 mag after interstellar dust correction, and on their GAIA G apparent brightness with G$<$13.5 mag. Then, for group F alone, we added a final cut to reject stars that presented magnitude limits in the [12] or [24] WISE photometric bands. After a cross match with the SIMBAD database, we selected 89 supplementary targets after removing two known RCB stars: \object{ASAS J050232-7218.9} and \object{UX Ant}.

\begin{table*}[!htbp]
\caption{Number of selected Galactic objects after each selection criterion\label{tab.Selection}}
\medskip
\centering
\begin{tabular}{lrrrrrc}
\hline
\hline
\multicolumn{1}{c}{Selection}  & \multicolumn{4}{c}{Groups} & Total &  Fraction of stars with  \\
\multicolumn{1}{c}{criterion} &   \multicolumn{1}{c}{A} & \multicolumn{1}{c}{B} & \multicolumn{1}{c}{C} & \multicolumn{1}{c}{D} &       & E(B-V)$\leqslant$0.5  \\ 
 & \multicolumn{2}{c}{Parallax S/N$\geq$5} & \multicolumn{2}{c}{3$\leq$Parallax S/N$<$5} & &  \\
  & \multicolumn{1}{c}{ALLWISE} & \multicolumn{1}{c}{2MASS} & \multicolumn{1}{c}{ALLWISE} & \multicolumn{1}{c}{2MASS} & &  \\
\hline
Cut \#0 (initial set)           & 1484627  & 1007019  & 509965  & 122681 & 3124292  &  $\sim$85\%    \\ 
Cut \#1         & 3218  &  891 & 720  & 488 & 5317 & $\sim$67\%    \\ 
Cut \#2         & 552   & 162  & 132 & 170 & 1016  &  $\sim$48\%    \\ 
Cut \#3         & 299   & 78  &  118 & 154 &  649 &  $\sim$45\%    \\ 
Known dLHdC/RCB  &   12$^{\star A}$ & 4$^{\star B}$  & 1$^{\star C}$ & 1$^{\star D}$ & 18  &  $\sim$55\%   \\ 
$\hspace{2 mm}$stars removed & \multicolumn{6}{c}{}\\
\hline
Total selected  &  287  & 74 & 117 & 154 & 631 &   $\sim$44\%    \\ 
Observed        & 139  & 43 & 84  & 91 & 357 &  $\sim$65\%    \\ 
New dLHdC/RCB   &  11   &  6  & 9  & 0  & 26  &  $\sim$46\%   \\ 
$\hspace{2 mm}$stars found & \multicolumn{6}{c}{}\\
\hline
\multicolumn{7}{l}{$^{\star A}$: \object{HD 137613}, \object{HD 173409}, \object{HD 148839}, \object{HD 175893}, \object{V2552 Oph}, \object{V532 Oph}, \object{[TCW2013] ASAS-RCB-10}, } \\
\multicolumn{7}{l}{\object{[TCW2013] ASAS-RCB-3}, \object{[TCW2013] ASAS-RCB-8}, \object{GU Sgr}, \object{WISE J172447.52-290418.6} and \object{WISE J172951.80-101715.9}} \\
\multicolumn{7}{l}{$^{\star B}$: \object{HD 182040}, \object{RT Nor}, \object{RY Sgr} and \object{SV Sge} ; $^{\star C}$: \object{WISE J182943.83-190246.2} ; $^{\star D}$: \object{U Aqr}} \\
\hline
\end{tabular}
\end{table*}

\section{Discoveries \label{sec_discoveries}}

\subsection{Spectroscopic follow-up}

The spectroscopic follow-up of all targets was conducted with the Wide Field Spectrograph (WiFeS) instrument \citep{2007Ap&SS.310..255D} attached to the 2.3 m telescope of the Australian National University at Siding Spring Observatory (SSO). WiFeS is an integral-field spectrograph permanently mounted at the Nasmyth A focus. It provides a $25\arcsec \times38\arcsec$~field of view with 0.5$\arcsec$ sampling along each of the twenty-five $38\arcsec \times1\arcsec$~slitlets. The visible wavelength interval is divided by a dichroic at around 600 nm, feeding two essentially similar spectrographs. The spectra have a two-pixel resolution of 2 $\AA$ and wide wavelength coverage, from 340 to 960 nm. We observed 427 targets with WiFeS during 21 nights between April and October 2021. The collected spectroscopic dataset was reduced using the PyWiFeS data reduction pipeline \citep{2014Ap&SS.349..617C}.

\begin{figure}
\centering
\includegraphics[width=3.5in,origin=c]{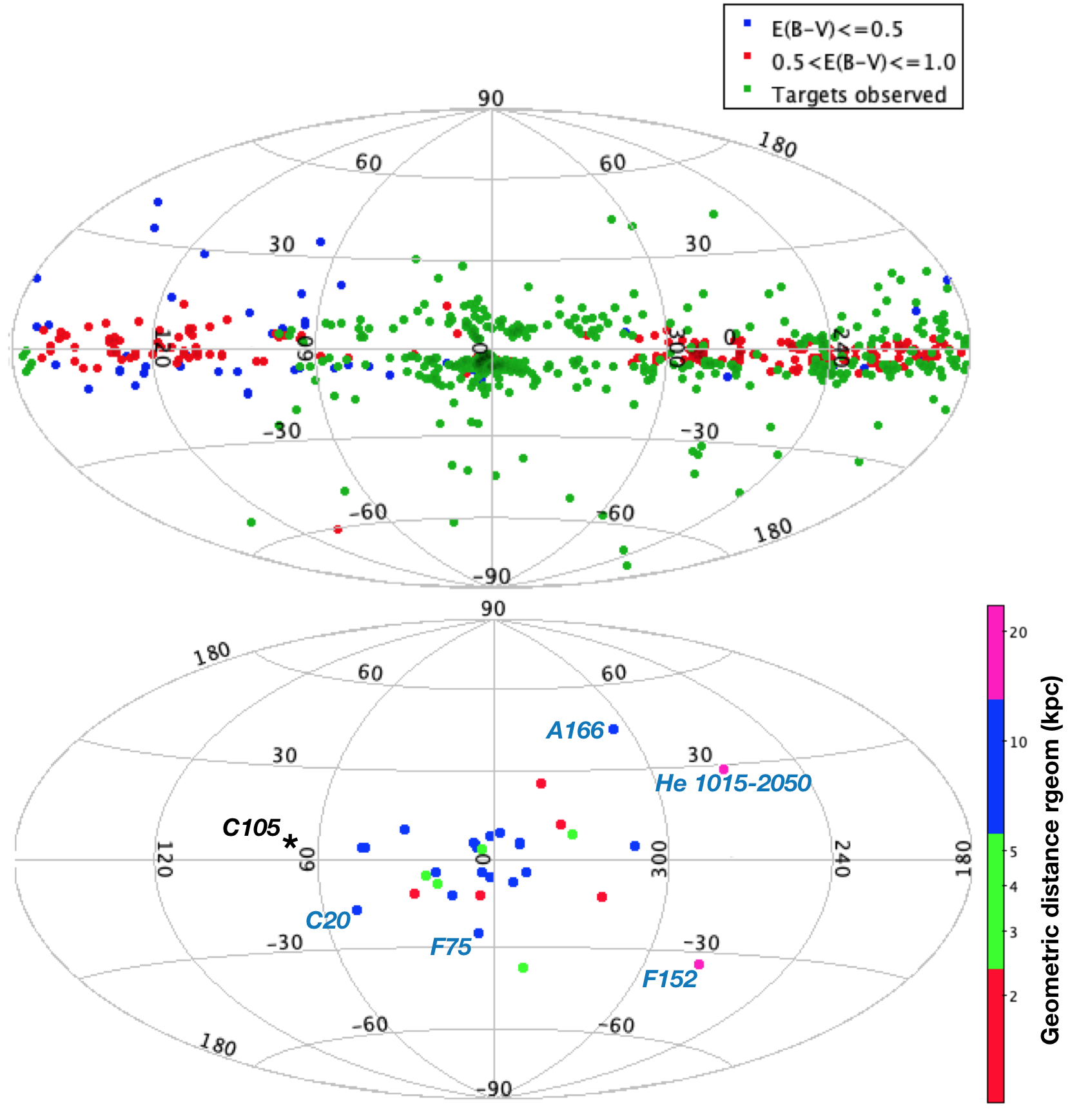}
\caption{Spatial distribution in Galactic coordinates. Top: All selected targets for spectroscopic follow-up with those already observed in green; the remaining targets are shown in blue for  an interstellar reddening lower than E(B-V)$\leqslant$0.5 mag, and red for higher reddening of up to 1.0 mag. Bottom: All dLHdC stars colour-coded with their respective geometric distances rgeom inferred by \citet{2021AJ....161..147B}. The names of the dLHdC stars that are almost certainly located in the Galactic halo are indicated. The location of the new RCB star, C105, is indicated with a black star.}
\label{fig_GalacticDistrib}
\end{figure}

The sky distribution of all targets selected for spectroscopic follow-up is presented in Figure~\ref{fig_GalacticDistrib}. Nearly $\sim$60\% of all selected targets were observed, which included almost all those located within 45 degrees of the Galactic centre. There, we had almost no available targets located within 3 degrees of the Galactic plane. Seventy-eight percent of the targets we did not observe (i.e. 31\% of total) are located in a sky area that could not be reached with the telescope during our seven-month-long observing campaign, that is, Dec$>$+25 deg and/or 6$<$RA$<$10 H. All the remaining unobserved targets were considered low-priority targets because at least one magnitude limit was reported in one of the three photometric 2MASS IR bands. Most of the time, this indicated that the source was strongly blended with a nearby object.

After inspection of the collected spectra, we found 27 new dustless HdC stars, one new RCB star, and two extreme helium (EHe) stars. They are listed in Table~\ref{tab.newHdC} with the names used in the study, their GAIA eDR3 parallax S/N and magnitudes, and their estimated geometric distance with a 1$\sigma$ error \citep{2021AJ....161..147B}. We first discuss the new HdC stars and then provide details of the two new EHe stars in Section ~\ref{sec_eHe}.

\begin{table*}[!htbp]
\caption{Newly discovered HdC and EHe stars
\label{tab.newHdC}}
\medskip
\centering
\begin{tabular}{lclccccccc}
\hline
Star  & 2MASS  & Other  & RA & Dec & GAIA & Distance$^\star$  &  GAIA   & GAIA & E(B-V)$^{\diamond}$  \\
 & Id & names  &  &  & parallax &  (kpc) with  & G  & BP-RP & mag \\
 &  &  & & & S/N & 1 sig error  &  mag & mag & \\
\hline
\multicolumn{10}{c}{}\\
\multicolumn{10}{c}{\emph{Known Galactic dLHdC stars}}\\
\hline
 & 15274831-2510101 & HD 137613 & 15:27:48.32   & -25:10:10.13 & 24.8 & 1.23$_{-0.05}^{+0.04}$ & 7.18 & 1.27 & 0.16  \\  
 & 16354579-6707366 & HD 148839 & 16:35:45.79 & -67:07:36.70 & 30.0 & 1.73$_{-0.05}^{+0.05}$ & 8.16 & 0.93 & 0.07 \\  
 & 18462663-3120321 & HD 173409 & 18:46:26.63 & -31:20:32.08 & 28.3 & 2.06$_{-0.07}^{+0.07}$ & 9.35 & 0.85 & 0.15 \\  
 & 19231008-1042113 & HD 182040 & 19:23:10.08 & -10:42:11.54 & 39.1 & 0.88$_{-0.02}^{+0.02}$ & 6.71 & 1.19  & 0.18 \\  
 & 10173423-2105138 & HE 1015-2050 & 10:17:34.23        & -21:05:13.88 & -0.65 & 12.00$_{-3.6}^{+11.4}$ & 15.92 & 1.02 & 0.04 \\  
\multicolumn{10}{c}{}\\
\multicolumn{10}{c}{\emph{New Galactic dLHdC stars}}\\
\hline
A166 & 13031356-1909225 &  & 13:03:13.56 & -19:09:22.68 & 6.94 & 6.20$_{-0.7}^{+0.8}$ & 12.03 & 1.91 & 0.08 \\  
A182 & 15473913-4435137 & \object{C* 2277}$^{r1}$ & 15:47:39.14 & -44:35:13.82 & 8.93 & 5.26$_{-0.4}^{+0.5}$ & 11.65 & 1.10 & 0.25 \\ 
A183 & 15503303-3944112 & SOPS IV e-67$^{r2}$ & 15:50:33.04 & -39:44:11.29 & 25.68 & 2.36$_{-0.1}^{+0.1}$ & 10.59 & 1.66 & 0.54 \\ 
A223 & 18563551-1609111 & \object{C* 2679}$^{r1}$ & 18:56:35.52 & -16:09:11.14 & 9.21 & 5.04$_{-0.4}^{+0.5}$ & 11.68 & 1.50 & 0.39 \\ 
A226 & 19022653-2228450 &  & 19:02:26.53 & -22:28:45.12 & 6.82 & 7.59$_{-0.6}^{+0.8}$ & 11.86 & 0.95 & 0.14\\  
A249 & 20393478-5123037 & \object{CD-51 12650}$^{r3}$ & 20:39:34.78 & -51:23:03.85 & 8.61 & 5.34$_{-0.4}^{+0.7}$ & 10.46 & 0.62 & 0.03 \\ 
A770 & 17024456-3335155 &  & 17:02:44.56 & -33:35:15.49 & 6.01 & 7.23$_{-0.9}^{+1.3}$ & 12.34 & 1.85 & 0.54 \\ 
A811 & 18541725-1150230 &  & 18:54:17.25 & -11:50:23.16 & 10.51 & 4.82$_{-0.4}^{+0.4}$ & 12.99 & 1.66 & 0.53 \\ 
A814 & 18592834+1149340 &  & 18:59:28.34 & +11:49:33.84 & 6.68 & 8.35$_{-1.1}^{+1.2}$ & 12.58 & 2.03 & 0.91 \\ 
A977 & 17424229-2418056  &   & 17:42:42.29  & -24:18:05.78   & 6.68 & 4.37$_{-0.5}^{+0.5}$  & 13.08  & 2.44  & 0.87  \\ 
A980 & 18113561+0154326  &  &  18:11:35.62 & +01:54:32.59 & 6.24 & 7.91$_{-1.0}^{+1.1}$  & 10.29  & 0.70  & 0.30 \\ 
B42 & 18134081-3129543 & \object{SOPS II e-12}$^{r2}$ & 18:13:40.81 & -31:29:54.32 & 5.92 & 7.39$_{-0.8}^{+1.1}$ & 11.54 & 1.71 & 0.38 \\ 
B563 & 17192903-2428337 &  & 17:19:29.03 & -24:28:33.88 & 8.38 & 5.69$_{-0.4}^{+0.5}$ & 12.93 & 2.03 & 0.80 \\ 
B564 & 17342620-4105220 &  & 17:34:26.19  & -41:05:22.17 & 7.73 & 6.70$_{-0.6}^{+0.8}$ & 11.98 & 1.02 & 0.59 \\ 
B565 & 17404848-2032199 &  & 17:40:48.48  & -20:32:20.15 & 5.28 & 6.90$_{-0.9}^{+0.9}$ & 12.97 & 1.57 & 0.72 \\ 
B566 & 17422375-2132006 &  & 17:42:23.74  & -21:32:00.74 & 8.15 & 6.25$_{-0.7}^{+0.9}$  & 12.40 & 1.73 & 0.61\\ 
B567 & 17430564-2224574 &  & 17:43:05.65  & -22:24:57.42 & 5.82 & 6.80$_{-0.7}^{+1.0}$ & 13.22 & 1.73 & 0.88 \\ 
C17 & 17054013-2616565 &  & 17:05:40.13 & -26:16:56.67 & 3.52 & 10.47$_{-1.4}^{+1.5}$ & 12.43 & 1.09 & 0.30 \\ 
C20 & 20194278+0504039 & \object{C* 2891}$^{r1}$ & 20:19:42.79 & +05:04:04.04 & 4.68 & 10.60$_{-1.2}^{+1.7}$ & 12.79 & 1.02 & 0.13 \\ 
C27 & 18013219-3856322 & \object{C* 2510}$^{r1}$ & 18:01:32.19 & -38:56:32.30 & 3.54 & 10.19$_{-1.2}^{+1.3}$ & 11.63 & 1.29 & 0.20 \\ 
C38 & 18114332-2814211 &  & 18:11:43.32  & -28:14:21.32       & 3.67 & 7.92$_{-1.0}^{+1.2}$  & 11.78 & 1.49 & 0.40 \\ 
C526 & 13564019-5740012 &  & 13:56:40.21 & -57:40:01.28 & 4.75 & 9.00$_{-1.0}^{+1.4}$ & 11.87 & 1.22 & 0.54 \\ 
C528 & 17004761-3330556 &  & 17:00:47.62  & -33:30:55.81 & 4.70 & 7.30$_{-0.8}^{+0.9}$ & 13.72 & 1.76 & 0.57 \\ 
C539 & 18444149-1415021 &  & 18:44:41.49 & -14:15:02.21 & 3.39 & 8.72$_{-1.1}^{+1.4}$ & 13.77 & 2.11 & 0.77 \\ 
C542 & 18571994+1059217 &  & 18:57:19.95 & +10:59:21.61 & 4.68 & 10.09$_{-1.2}^{+1.7}$ & 12.75 & 1.92 & 0.80 \\ 
F75 & 19444060-3454422 & SOPS IIc- 65$^{r2}$ & 19:44:40.59 &  -34:54:42.33      & 1.98  & 11.28$_{-1.6}^{+2.3}$  & 11.31 & 0.80 & 0.18 \\ 
F152 & 05152898-6648374 &  & 05:15:28.99 & -66:48:37.42 & 0.90 & 23.61$_{-4.0}^{+5.0}$  & 12.46 & 0.79 & 0.14 \\ 
\hline
\multicolumn{10}{c}{}\\
\multicolumn{10}{c}{\emph{New Galactic RCB star}}\\
\hline
C105 & 19393548+3434471 &  & 19:39:35.49 & +34:34:47.19 & 3.50 & 11.72$_{-1.5}^{+2.3}$ & 11.23 & 1.13 & 0.23 \\ 

\hline
\multicolumn{10}{c}{}\\
\multicolumn{10}{c}{\emph{New Galactic EHe stars}}\\
\hline
A208 & 18244794-2214291 &  & 18:24:47.95 & -22:14:29.14 & 7.96 & 5.15$_{-0.4}^{+0.5}$ & 11.57 & 0.66 & 0.69 \\ 
A798 & 18335703+0529170 &  & 18:33:57.03  & +05:29:16.81 & 5.63 & 7.26$_{-0.7}^{+1.0}$  & 12.42 & 1.13 & 0.72 \\ 
\hline
\multicolumn{10}{l}{$^\star$: Geometric distances inferred by \citet{2021AJ....161..147B}; $^\diamond$: from \citet{2011ApJ...737..103S}}\\  
\multicolumn{10}{l}{$^{r1}$: \citet{1973PW&SO...1.....S,2001BaltA..10....1A}; $^{r2}$: \citet{1972PDAUC...2...59S}; $^{r3}$: The "Cordoba Durchmusterung" catalogue (1932)} \\

\end{tabular}
\end{table*}

\begin{figure*}
\centering
\includegraphics[width=7in]{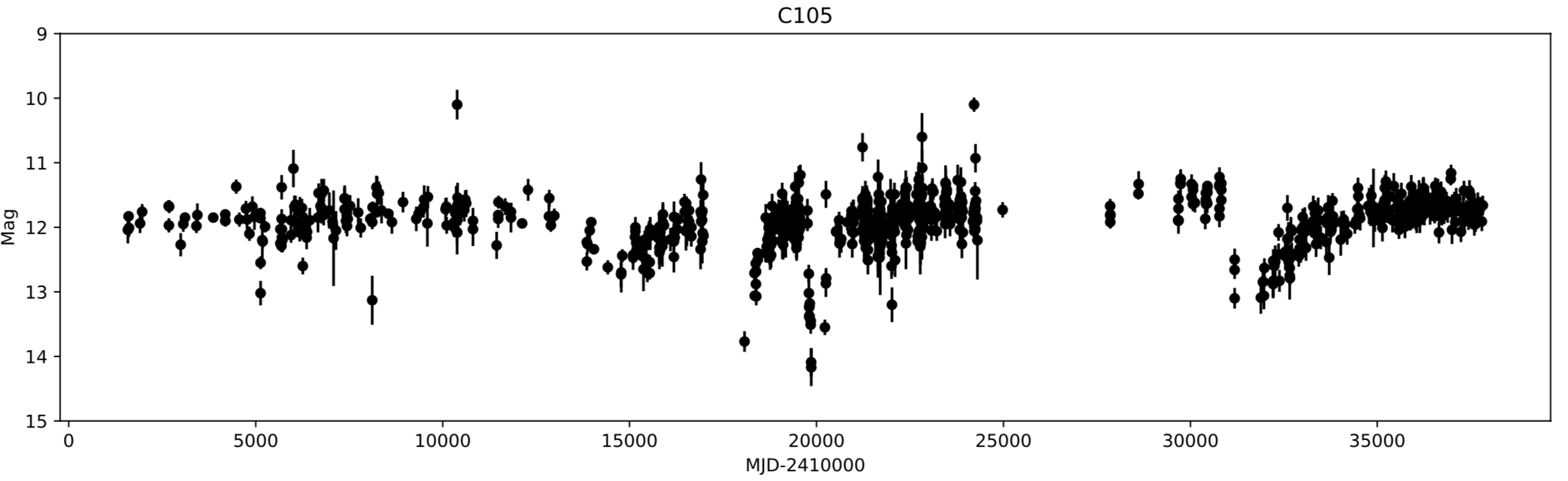}
\caption{DASCH light curve of C105 covering about 100 years from $\sim$1890 to $\sim$1990. The average plate limit is $\sim$13.5 mag.}
\label{fig_lc_C105}
\end{figure*}

\begin{table}[]
\caption{Spectroscopic and photometric characteristics observed in new and known HdC stars
\label{tab.spectroHdC}}
\medskip
\centering
\begin{tabular}{clllcl}
\hline
stars  & \multicolumn{2}{c}{Line strength}  & Temp. & M$_V$ & (V-I)$_0$  \\
  &  H &  Li & group & (mag) & (mag) \\
\hline
\multicolumn{6}{c}{\emph{dLHdC stars}}\\
\hline
HD 137613 &  &   & Mild & -3.29$_{-0.08}^{+0.09}$  &  0.89  \\
HD 148839 & ++ & +  & Mild & -3.09$_{-0.07}^{+0.06}$ &  0.66  \\
HD 173409 &  &   & Mild & -2.47$_{-0.07}^{+0.07}$ &  0.68  \\
HD 182040 &  &   & Cold & -3.28$_{-0.05}^{+0.05}$ &  0.74  \\
HE 1015-2050 &  &   & Cold & nd &  0.88 \\
\hline
A166 & &  & Cold & -1.38$_{-0.27}^{+0.27}$ & 1.80  \\
A182 & & & Warm & -2.46$_{-0.20}^{+0.19}$ & 0.63$^{\bigtriangleup}$  \\ 
A183 & ++ & + & Mild & -2.37$_{-0.09}^{+0.07}$ &  0.84 \\ 
A223 & + & & Cold & -2.96$_{-0.19}^{+0.16}$ &  0.70 \\ 
A226 & & & Mild & -2.85$_{-0.22}^{+0.19}$ &  0.60 \\
A249 & ++ &  & Warm & -3.22$_{-0.26}^{+0.18}$ & 0.45 \\ 
A770 & + & & Cold & -2.87$_{-0.36}^{+0.27}$ & 1.10 \\
A811 & ++ & + & Cold & -2.05$_{-0.18}^{+0.18}$ & 0.62 \\ 
A814 & ++ &  & Mild  & -4.26$_{-0.29}^{+0.30}$  & 0.64 \\
A977 &  &  & Cold & -2.75$_{-0.22}^{+0.27}$ & 0.73  \\
A980 & + & + &  Warm & -5.18$_{-0.29}^{+0.28}$ & 0.16$^{\bigtriangledown}$ \\
B42 & + & & Mild & -3.39$_{-0.29}^{+0.24}$ & 0.96  \\ 
B563 & + & ++ & Mild & -2.96$_{-0.20}^{+0.18}$ & 0.75  \\
B564 & + & + & Warm & -3.75$_{-0.24}^{+0.19}$ & 0.11$^{\bigtriangledown}$  \\
B565 & ++ & + & Mild & -2.97$_{-0.26}^{+0.30}$ &  0.57 \\
B566 &  & & Mild & -3.07$_{-0.28}^{+0.25}$ & 0.78  \\
B567 & & & Warm & -3.33$_{-0.30}^{+0.25}$ & 0.27 \\
C17 & ++ & +  & Mild & -3.55$_{-0.29}^{+0.32}$  & 0.47$^{\bigtriangledown}$  \\
C20   & + & + & Mild & -2.41$_{-0.32}^{+0.26}$ &  0.82 \\ 
C27 & & & Cold & -3.65$_{-0.26}^{+0.27}$ & 0.85  \\ 
C38   & + & + & Mild & -3.71$_{-0.30}^{+0.31}$ & 0.76  \\
C526  & + &  & Warm & -4.29$_{-0.31}^{+0.26}$ & 0.35  \\ 
C528 & ++ & + & Mild & -1.81$_{-0.24}^{+0.26}$ &  0.81 \\
C539  & & & Mild & -2.98$_{-0.33}^{+0.31}$ & 0.84  \\
C542 & ++ & + & Mild & -4.14$_{-0.34}^{+0.29}$ & 0.70  \\
F75   & + & + & Warm & -4.41$_{-0.40}^{+0.32}$ &  0.39 \\
F152 & Em. &  & Warm & -4.76$_{-0.42}^{+0.40}$ & 0.47  \\ 
\hline
\multicolumn{6}{c}{\emph{RCB stars}}\\
\hline
C105 & & & Mild & -4.64$_{-0.39}^{+0.29}$  & 0.61  \\ 
HD 175893  & & & Mild & -3.99$_{-0.15}^{+0.15}$ &  0.61  \\ 
\hline
\multicolumn{6}{l}{$^{\bigtriangleup}$Interstellar reddening correction possibly underestimated}\\
\multicolumn{6}{l}{$^{\bigtriangledown}$Interstellar reddening correction possibly overestimated}\\
\end{tabular}
\end{table}

\subsection{New dLHdC and RCB stars \label{sec_newHdC}}

The spectra of all new HdC stars stand out. We principally searched for an absence or weakness in any hydrogen Balmer lines and the absence of the CH band head located at around $\sim$4300 \AA. Then we individually compared the selected spectra to those of the 130 known HdC stars we collected so far to recognise the typical metallic and carbon absorption patterns in relation to the effective temperature, that is, the C I absorption lines for the warm HdC stars (T$_{eff}>$6500 K) and the CN and C$_{2}$ band heads observed for the colder stars. New HdCs were required to have spectra closely resembling at least one of the known HdC stars in order to be confirmed as members of the class. Detailed discussions of the characteristics of these spectra and some presentations of them can be found in \citet{2020A&A...635A..14T}. Additionally, we note that 6 of the new dLHdC stars (A182, A183, A223, B42, C20, and C27) were already catalogued as carbon-rich stars in the general catalogue of carbon stars \citep{2001BaltA..10....1A}. A spectral classification of type R (R2 for A182) was given by the CDS/SIMBAD database, indicating they are warmer than classical carbon stars. 

One first marked difference we observed is the general weakness of the CN bands in the new dLHdC stars compared to known RCB stars. For a new dLHdC star and a known RCB star with similar C$_2$ band strengths, and thus presumably similar temperatures, the dLHdC stars generally show significantly weaker CN bands. Assuming the similarities in C$_2$ band strength of these two stars indicates a similar carbon abundance, the weakness of the CN bands implies that the new dLHdC stars have a lower nitrogen abundance. More details of this comparison will be made available by Crawford et al. (2022, in prep.). However, we note that this lower N abundance was not detected in the \citet{2009ApJ...696.1733G} high-resolution IR spectroscopic analysis. They observed  3 of the known dLHdC stars and three RCB stars (HD 175893, S Aps, and Y Mus), and reported similar high N abundances for all. Nonetheless, however, we should be cautious with their result as they commented that their derived N abundances for the dLHdC star, HD 137613, and the RCB star, Y Mus, agree only poorly (difference of $\sim$0.7 dex) with other analyses made by \citet{2002BaltA..11..249K} and \citet{2000A&A...353..287A}, respectively, based on high-resolution visible spectroscopy, and that the issue was not properly investigated. Because the abundance of most metals, such as N, are roughly proportional to the abundace of Fe \citep{2011MNRAS.414.3599J} in most RCB stars, we can infer that the metallicity of dLHdC stars is probably lower than in most HdC stars.  Future investigations are required to confirm this.

\begin{figure*}
\centering
\includegraphics[width=6in]{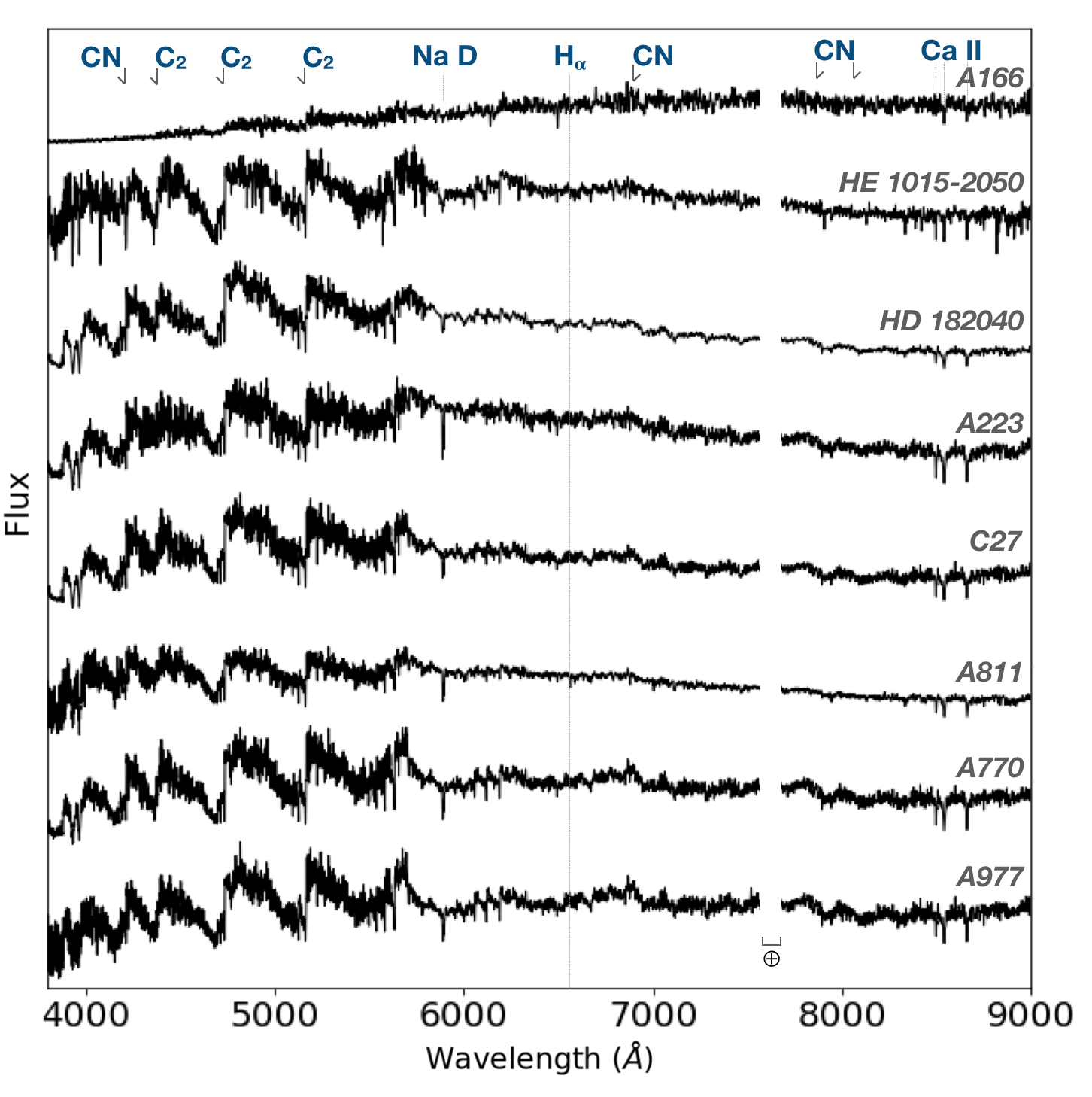}
\caption{Spectra from 3800 to 9000 $\AA$ of the dLHdC stars we consider as cold, T$_{eff}\leq$ 5500 K. The elements related to the more prominent absorption lines and the expected position of H$_\alpha$ are indicated with vertical dashed lines, as are various band heads of C$_2$ and CN. The gaps in the spectra corresponding to telluric lines are indicated with a crossed circle. The names of the corresponding stars are given on the right side. The ordinate is arbitrary.}
\label{fig_Spectra_newHdC_cool}
\end{figure*}
   
\begin{figure*}
\centering
\includegraphics[width=6in]{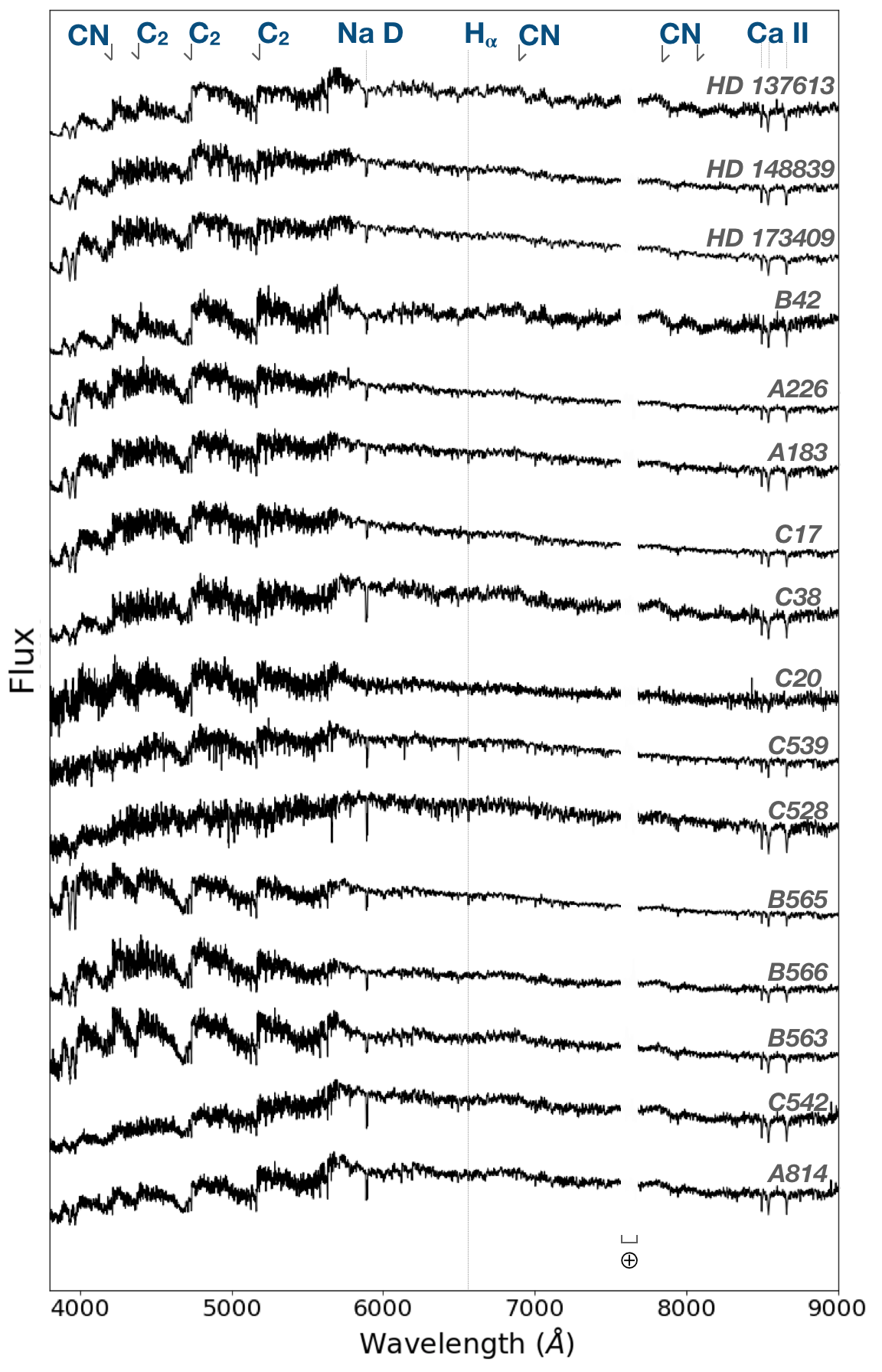}
\caption{Same as Figure~\ref{fig_Spectra_newHdC_cool}, but for dLHdC stars we considered to have a mild temperature, T$_{eff}$ between $\sim$6000 and $\sim$7000 K.}
\label{fig_Spectra_newHdC_mild}
\end{figure*}

\begin{figure*}
\centering
\includegraphics[width=6in]{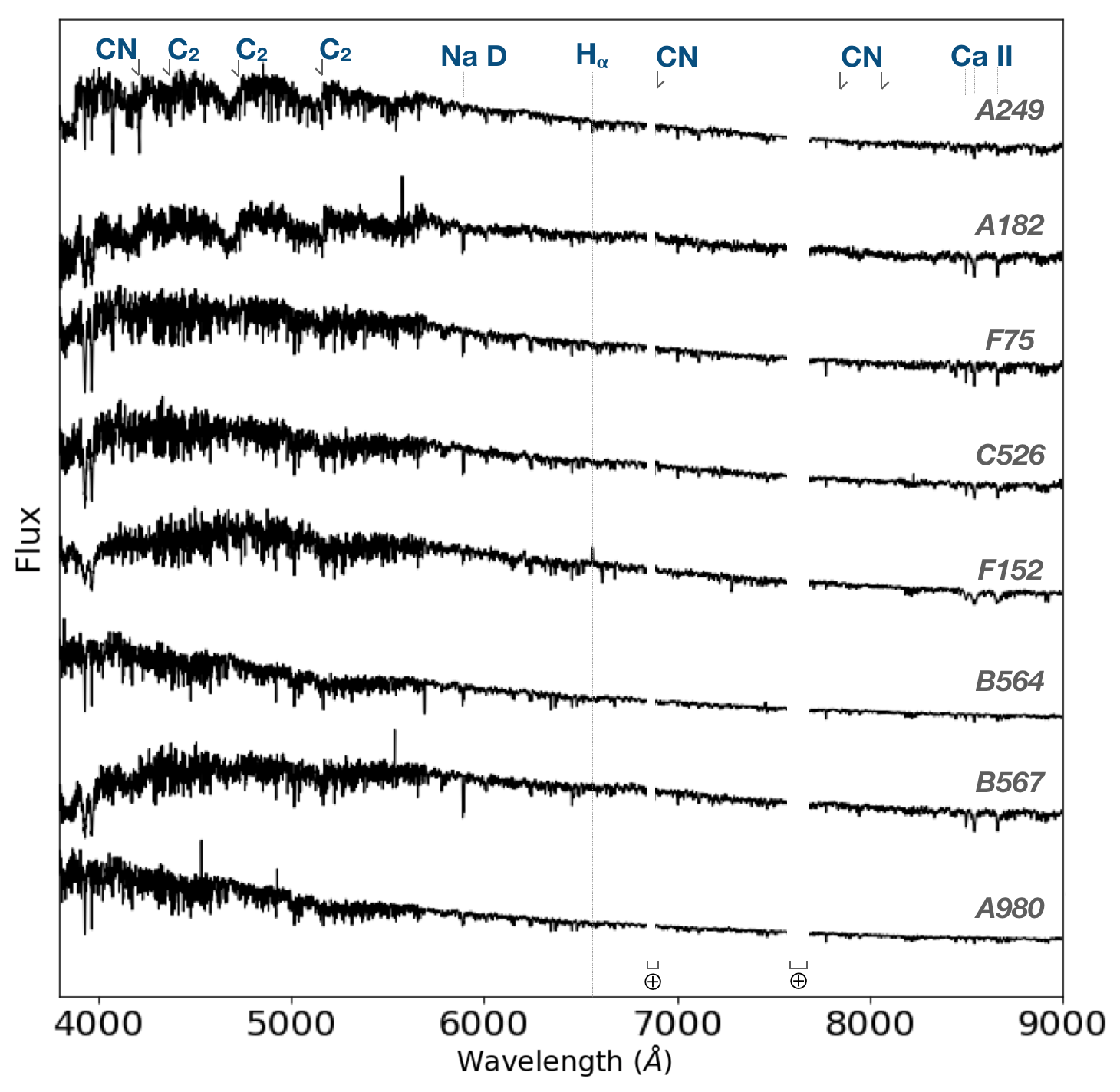}
\caption{Same as Figure~\ref{fig_Spectra_newHdC_cool}, but for dLHdC stars we considered to have a warm temperature, T$_{eff}$ between $\sim$7000 and $\sim$8000 K.}
\label{fig_Spectra_newHdC_warm}
\end{figure*}

\begin{figure*}
\centering
\includegraphics[width=6in]{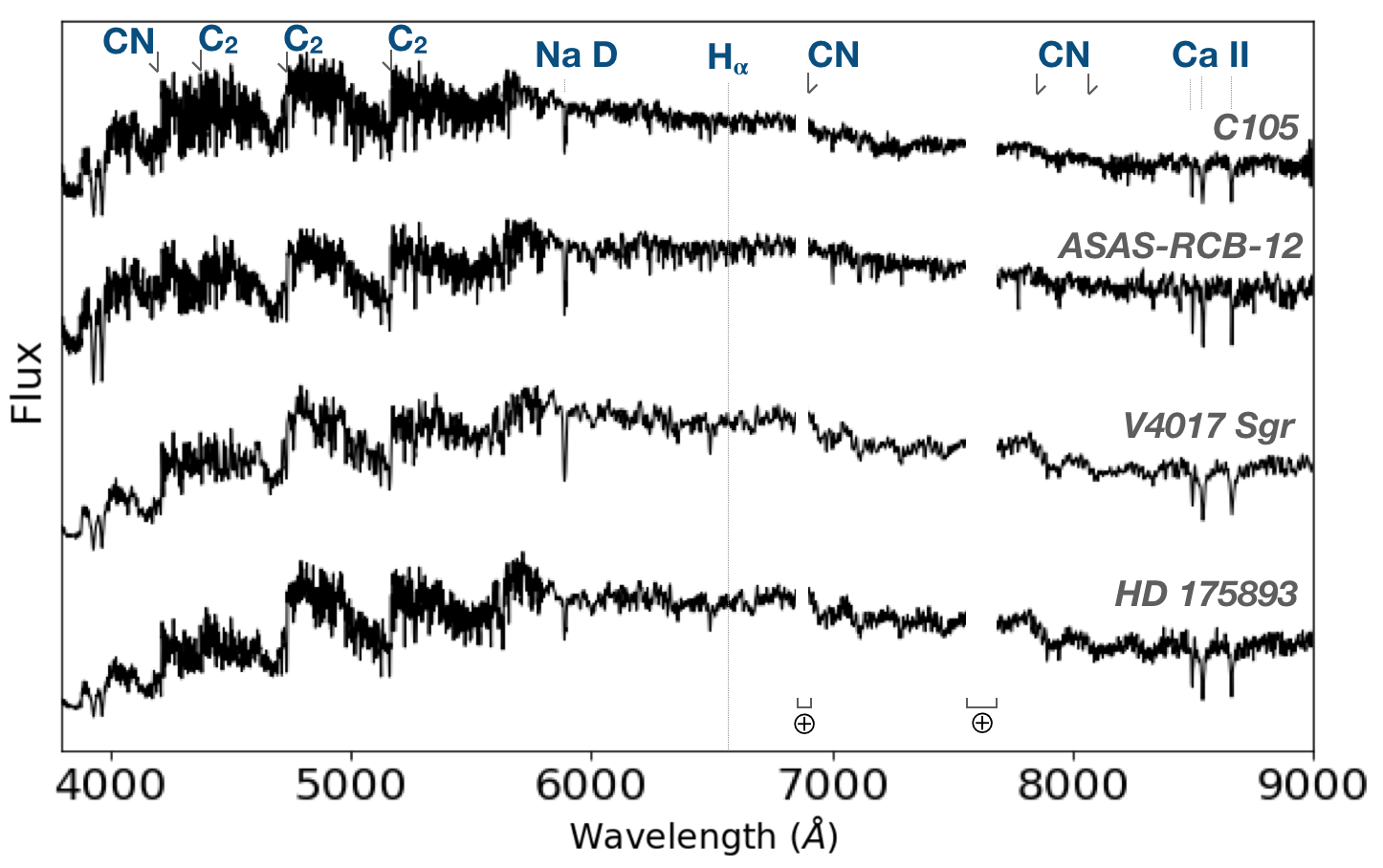}
\caption{Spectra from 3800 to 9000 $\AA$ of the newly discovered RCB star, C105, and HD 175893, which was considered for a long time as a dLHdC star, but which is in fact surrounded by a warm circumstellar dust shell like classical RCB stars. They are compared to known RCB stars whose spectra are very similar. The names of the corresponding stars are given on the right side. The ordinate is arbitrary.}
\label{fig_Spectra_newRCB}
\end{figure*}

\begin{figure*}
\centering
\includegraphics[width=6in]{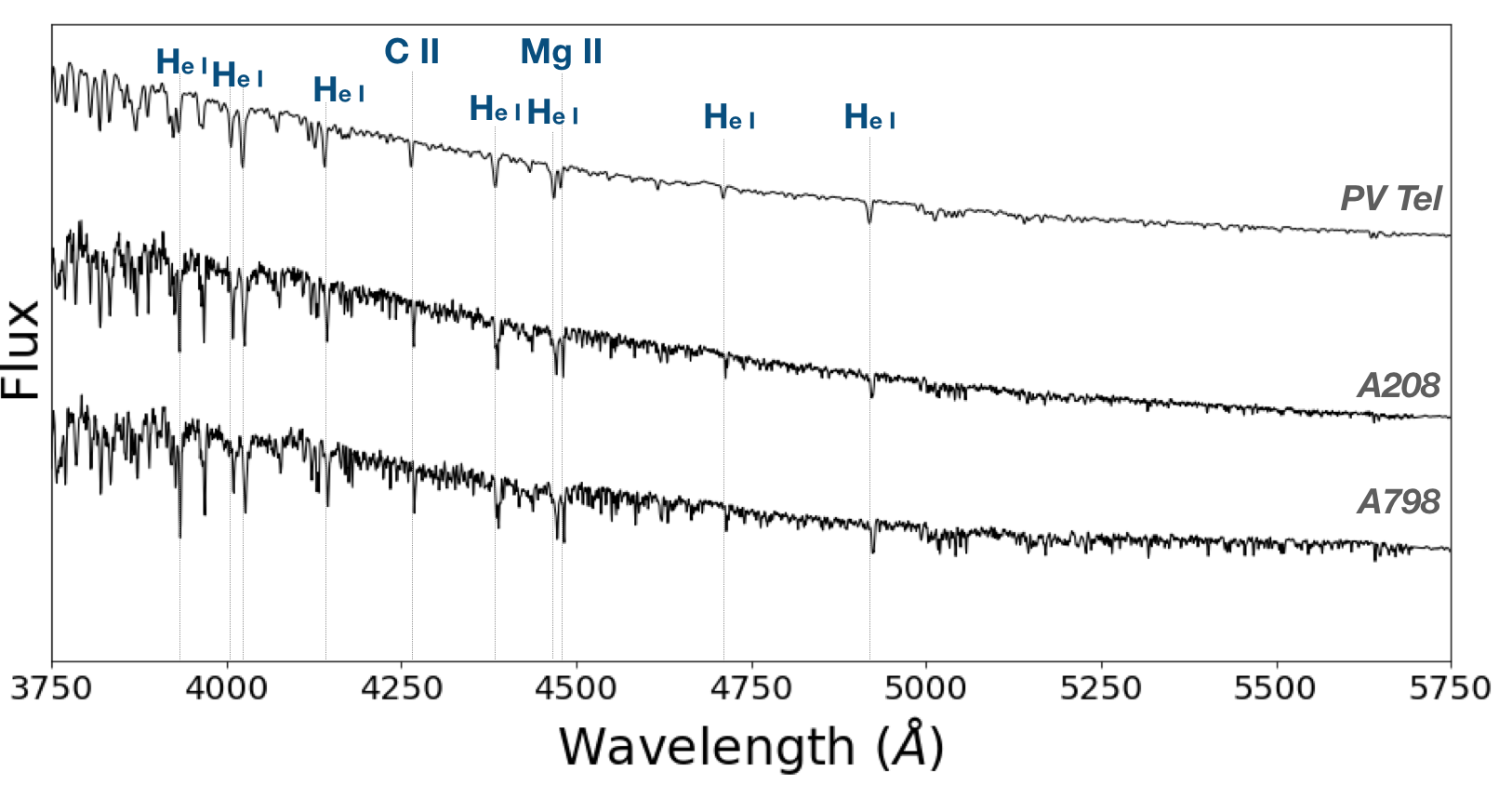}
\caption{Spectra in the blue region from 3750 to 5750 $\AA$ of the newly discovered EHe stars and the known EHe star \object{PV Tel}, which has a similar absolute brightness and temperature. The names of the corresponding stars are given on the right side. The ordinate is arbitrary.}
\label{fig_Spectra_eHe}
\end{figure*}

From a careful spectral comparison with known HdC stars and models, we also attributed a temperature to all new dLHdC stars, from the coldest, A166 (T$_{eff}\sim$4000 K), to the warmest, B564 (T$_{eff}\sim$8000 K). A detailed study of this work and for the 130 known RCB stars can be found in Crawford et al. (2022, in prep.). We then classified them into the three following groups depending on temperature range: cold for T$_{eff}\leq$5750 K, mild for 5750$<$T$_{eff}\leq$6500 K, and warm for 6500$<$T$_{eff}\leq$8000 K. These groups are listed in Table~\ref{tab.spectroHdC}, and their spectra are presented in Figures ~\ref{fig_Spectra_newHdC_cool} -- ~\ref{fig_Spectra_newHdC_warm}. We found that the majority of the new dLHdC stars (13) have a temperature similar to the 3 historically known stars and belong to the mild group, while 8 are classified as warm and the remaining 6 are considered as cold, of which 5 have a temperature near 5500 K. We were surprised to discover warm dLHdC stars as none was known before and our analysis did not target them specifically. Their temperatures are similar to those of Y Mus and V3795 Sgr. They were all discovered, except for A249, thanks to the interstellar extinction that reddened their BP-RP colour above the 0.5 mag threshold we used to create the initial star list (Section~\ref{sec_starlist}). We will widen up our search to bluer stars in future surveys.

The hydrogen deficiency is real in all newly discovered dLHdC stars. This is also confirmed by their near-IR spectra as reported by \citet{Karambelkar_2022}. However, surprisingly, the hydrogen deficiency of many of the new dLHdC stars is somewhat lower than that of most known HdC stars. We found traces of H$_\alpha$ in 19 out of the 27: 3 belong to the cold group ($\sim$50\%), 10 belong to the mild group ($\sim$77\%), and 6 to the warm group ($\sim$66\%). They are listed in Table~\ref{tab.spectroHdC} with a scale from 0 to 2 "+" indicating the strength of the observed H$_\alpha$ absorption line (see also Fig.~\ref{fig_Spectra_newHdC_cool} to ~\ref{fig_Spectra_newHdC_warm} for the spectra). One of the five previously known dLHdC stars, HD 148839, presented this characteristic before. See \citet{1967MNRAS.137..119W} for a detailed discussion. Previously, abundance analyses of RCB and EHe stars have shown an apparent anti-correlation between the H and Fe abundances \citep[Fig.7]{2000A&A...353..287A}. Here, our observations of H and N abundances for dLHdC stars go towards that trend.  We even found a small emission H$_\alpha$ line in the spectrum of the warm HdC star, F152, as well as some emission in the Ca II H\&K and triplet lines. RCB stars generally show weak or absent Balmer lines, but at least one of them, V854 Cen, shows significant hydrogen lines \citep{1989MNRAS.240..689L,1989MNRAS.238P...1K}. \citet{Lawson_1992} even detected H$_\alpha$ in emission during a decline event. Recently, \citet{2020A&A...635A..14T} reported a similar spectroscopic behaviour in a new RCB star located in the Large Magellanic Cloud, called WISE\_J054221.91-690259.3 (or MSX-LMC-1795). However, for F152, the H$_\alpha$ emission is not correlated with a detected decline event as no such episode was seen in the F152 ASAS-SN light curve at JD$\sim$2459452.3 (25 August 2021). At this time, F152 was even at a peak of its irregular photometric variability. Furthermore, we note that F75, another warm new HdC star, was also reported in the past to show some degree of emission lines, as \citet{1972PDAUC...2...59S} classified it as "Ge pec" for its spectral type, which was therefore already seen as peculiar.

We also searched for traces of lithium in our medium-resolution spectra using the Li I 6707 absorption line. We found one strong occurrence in the spectrum of B563 and some indications for the possible presence of this line in the spectra of 11 others (see Table~\ref{tab.spectroHdC}) that would be of comparable strength to the one seen in HD 148839. They all need to be confirmed with higher-resolution spectroscopy. Interestingly, the new HdC star B565 (for which we observed a small absorption line at the location expected for Li I 6707) has been observed by the GALAH survey \citep{2021MNRAS.506..150B}, which  uses the multi-fiber 2df/HERMES high-resolution (R$\sim$28 000) spectrograph (Object Id 170508006401160). The GALAH survey published a complete abundance analysis for this star and confirmed the high abundance in B565 of carbon and oxygen, as well as of lithium. Furthermore, their abundance analysis reveals that B565 is a complete outlier in a diagram showing the abundance pattern of iron versus $\alpha$-process elements (with [$\alpha$/Fe]$\sim$0.99 and [Fe/H]$\sim$-0.82 for B565) compared to the nearly 700000 stars they followed-up.


Most RCB star atmospheres are known to have a low $^{13}$C/$^{12}$C isotopic ratio, but this is not always the case \citep{2008MNRAS.384..477R,2012ApJ...747..102H}. We searched for the presence of $^{13}$C in the new dLHdC stars spectra, particularly in the blue region, by comparing the $^{13}$C$^{12}$C absorption line located at 4744 $\AA$ and the nearby (1,0) $^{12}$C$^{12}$C at 4737 $\AA$. It was difficult to carry out this analysis in the red region reliably as the CN band heads are weak. We found strong evidence for weak $^{13}$C features in the spectra of A249, F75, or C526, but also possibly in the spectra of A166.

Among all the newly discovered HdC stars, we found one new RCB star, C105. Its spectrum is presented in Figure~\ref{fig_Spectra_newRCB} and is compared to that of \object{[TCW2013] ASAS-RCB-12}, an RCB star of similar temperature (i.e. T$_{eff}\sim$6000 K). They both show strong C$_2$ features, but some weak ones due to CN. This is in contrast with the two other spectra we present  from HD 175893 and V4017 Sgr, two RCB stars that also have a similar temperature. This indicates a lower nitrogen abundance in the atmosphere of C105 compared to more classical RCB stars. This was already seen in the atmosphere of all new dLHdC stars. C105 is a new RCB star because of the warm circumstellar dust we found to be surrounding it (see Fig.~\ref{fig_W2W3vsW3W4} and the discussion in Sections~\ref{sec_variability} and ~\ref{sec_dust}). However, we note that at the time of the 2MASS epoch (in 1998), the C105 \textit{JHK} magnitudes were not impacted by a warm thick dust shell as was detected during the WISE survey in 2010. Its colour indices, (J-H)$_0\sim$0.14 and (H-K)$_0\sim$0.13 mag, are similar to those of dLHdC stars that show no IR excess (see Fig.~\ref{fig_Select-2MASS}). This also explains why C105 was not selected in the all-sky RCB star candidates list made by \citet{2020A&A...635A..14T}. C105 was in a phase of low dust production in 1998. We also found a light curve from the DASCH\footnote{DASCH: Digital Access to a Sky Century at Harvard} project covering most of the past century, showing that C105 underwent some characteristic decline phases at least four times, with two major phases observed in 1940 and 1972 (Fig.~\ref{fig_lc_C105}).

It is not surprising to find new RCB stars as 16 of the known ones (14 in groups A to D, and 2 in groups F and G) have passed our selection cuts. These RCB stars did not show a high near-IR excess as no warm thick dust influenced the JHK photometry at the time of the 2MASS epochs.

\subsection{New extreme helium stars \label{sec_eHe}}

Extreme helium (EHe) stars are blue supergiant stars with surface temperatures ranging from 9000 to 35000 K. They are thought to share an evolutionary connection with HdC stars as the subsequent phase of the latter as they share similar spectroscopic characteristics due to an atmosphere that is poor in hydrogen, but highly enriched in carbon \citep{2008ASPC..391...53J,2011MNRAS.414.3599J,2001MNRAS.324..937P}.

Most known EHe stars (13 out of 21) were not part of the initial datasets of GAIA giant stars used in our analysis because their uncorrected BP-RP colour indexes were bluer than 0.5 mag. Furthermore, of the remaining 8 EHe stars, none passed the criteria we applied to the photometry recorded in the 2MASS and GAIA eDR3 catalogues. It is therefore surprising that we discovered 2 new EHe stars as they were not an objective of our search. Their discovery is due to a fortunate error. A mistake in the process of calculating the E(B-V) value was made for the stars belonging to group A, which resulted in reddening values related to sky areas located between 20 to 40 arcmins away from each star. Therefore, some stars were observed spectroscopically while they should have been left out of the selection. The new EHe stars, A208 and A798, are two of them. From the distribution of these stars, the error is the equivalent of having randomly observed some stars located on the bluer side of the J-H versus H-K colour-colour diagram, within 0.05 mag from the selection limit. These stars are distributed homogeneously in the GAIA M$_{G}$ versus (BP-RP)$_{0}$ colour-magnitude diagram with a (BP-RP)$_{0}$ colour ranging between -1.5 and 1.0 mag. Of the 656 stars originally selected in group A with the wrong E(B-V) reddening estimate, 333 should not have been selected. Of this sample, 203 had no spectroscopic type defined in CDS/SIMBAD and were therefore confirmed targets for spectroscopic follow-up. We observed 51 of them before correcting the mistake, and by good fortune, we found 2 new EHe stars, whose coordinates and geometric distances are listed in Table~\ref{tab.newHdC}.

The blue sides of the spectra are presented in Fig.~\ref{fig_Spectra_eHe} with a direct comparison with the known EHe star, \object{PV Tel}. Most of the strong absorption lines that are detected are due to He I, but some are also due to C I and Mg II. Many more lines can be identified using the list published by \citet{1981A&AS...44..349L} after studying a PV Tel high-resolution spectrum in detail. We calculated the visual absolute magnitudes and intrinsic colours (M$_V\sim$-3.60$_{-0.21}^{+0.17}$ and $\sim$-4.16$_{-0.27}^{+0.21}$ mag, (V-I)$_0\sim$-0.17 and $\sim$-0.04 mag for A208 and A798, respectively) and compared them to other EHe stars and HdC stars (see Fig.~\ref{fig_MV_VI}). They are located in a part of the HR diagram in which we find other cool EHe stars, such as \object{V2244 Oph}, \object{PV Tel}, \object{NO Ser}, \object{LSS 99}, \object{V1920 Cyg}, and \object{V4732 Sgr}.

Finally, we checked the light curves of both new EHe stars resulting from the ASAS-SN survey. We did not find any variability at the ASAS-SN photometric resolution for A208, but report some variabilities with a peak-to-peak amplitude of $\sim$0.1 mag and a periodicity of about 10 days for A798. This is confirmed by the monitoring by the CoRoT survey \citep{2014yCat....102028C}. The high-cadence light curve of A798, published with those of other stars observed in the faint-star mode (monitoring between July and September 2011), shows well-sampled irregular photometric variations, with a timescale between 5 to 10 days.

\section{Results and discussion \label{sec_result}}

\subsection{Galactic distribution of the new HdC stars \label{sec_distrib}}

The Galactic spatial distribution of all dLHdC stars is presented in Fig.~\ref{fig_GalacticDistrib}. Five of them (C20, F75, F152, A166, and HE 1015-2050) are almost certainly located in the Galactic halo, but most of the others are located around or inside the Galactic bulge, with geometric distances ranging between 6 and 12 kpc. We observed 216 candidates located around the Galactic centre (-45<\textit{l}<45 deg and |\textit{b}|<15 deg) and found 21 dLHdC stars, while similarly, we observed 135 candidates located along the Galactic disk (180<\textit{l}<300 deg and |\textit{b}|<15 deg) and found none. This is a clear indication that the dLHdC stars belong to the bulge population, implying that it is an old population group of stars, as expected from their hydrogen-deficient nature. Their sky distribution is similar to that of RCB stars \citep[Fig.16]{2020A&A...635A..14T}.

We note that F152 is located on the line of sight of the Large Magellanic Cloud, half-way to the largest satellite of the Milky\ Way, with a preferred median geometric distance of $\sim$24 kpc. In all likelihood, F152 cannot be located within the LMC as its absolute brightness would be 1.5 mag brighter than the brightest HdC stars ever observed (see Section~\ref{sec_cmd}).

\begin{figure*}
\centering
\includegraphics[width=5.3in,origin=c]{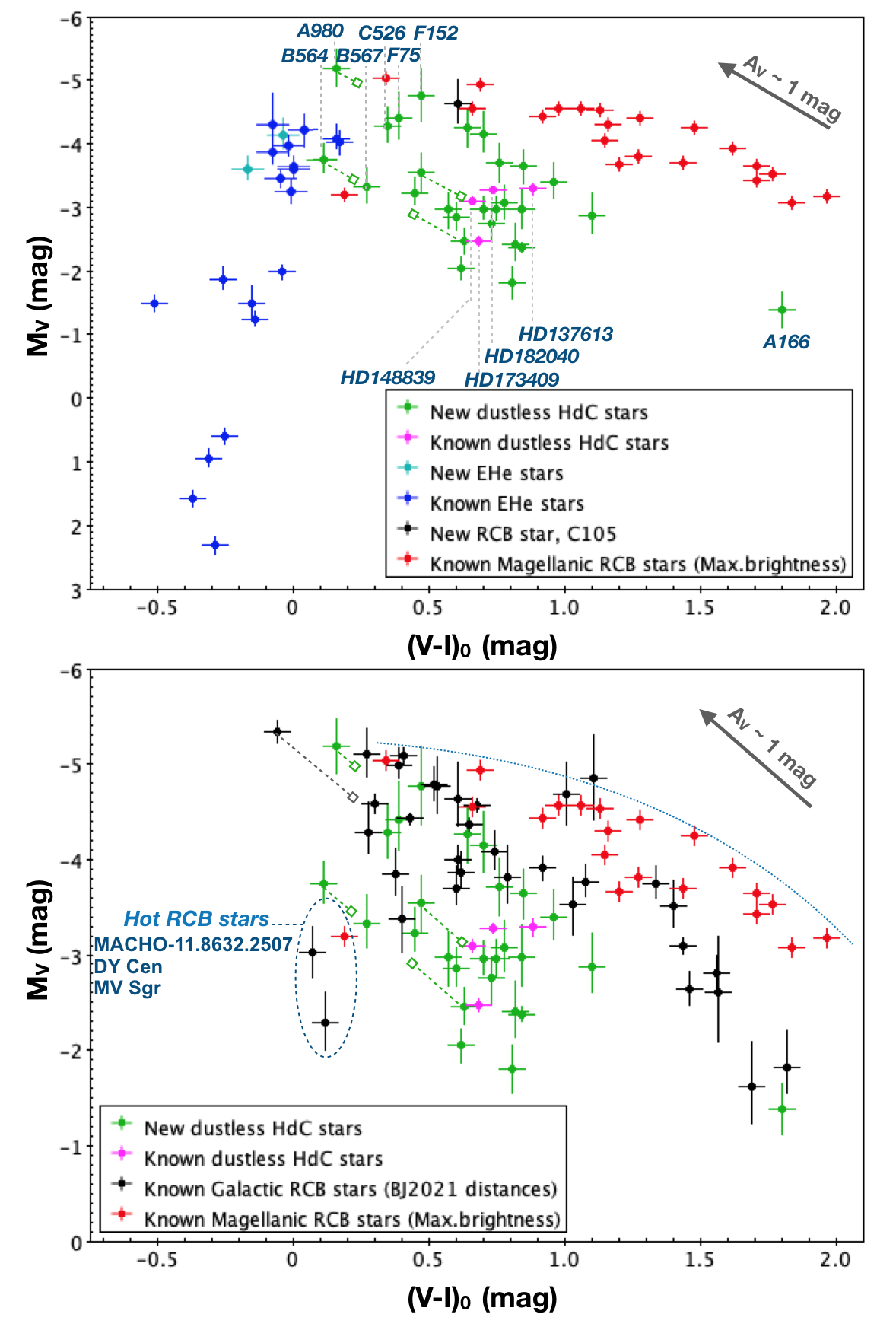}
\caption{Colour-magnitude diagrams M$_V$ vs (V-I)$_0$ with all dLHdC stars represented as well as most known EHe stars, some Magellanic RCB stars, and the Galactic RCB stars for which a GAIA eDR3 parallax was reported with S/N$>$2 and for which we were able to convincingly measure their maximum brightness in the V and I photometric bands. For all Galactic objects, we used the median geometric distances inferred by \citet{2021AJ....161..147B}, and the error bars on the luminosity scale reflect their associated 68\% confidence interval. Top: All stars except for the known Galactic RCB stars are represented. Bottom: Zoom on the brightest part of the diagram without the EHe stars and with the Galactic RCB stars. An interstellar reddening arrow representing an extinction of 1 mag in V is represented in both figures. The dashed lines, parallel to the extinction arrows, that are each associated with a warm RCB star (Y Mus), two warm dLHdC stars (A980 and B564), and two mild dLHdC stars (C17 and A182), correspond to the reddening over- and under-corrections we consider to have applied after estimating the respective temperatures from the stellar spectra. The locations of the hot RCB stars \object{MACHO 11.8632.2507}, \object{DY Cen}, and \object{MV Sgr} are indicated with an ellipse. The dotted curved indicates the maximum absolute M$_V$ magnitudes recorded for each bin of the (V-I)$_0$ colour index.}
\label{fig_MV_VI}
\end{figure*}

\subsection{Absolute brightness of the new HdC stars \label{sec_cmd}}

All newly discovered dLHdC stars range in apparent G brightness between $\sim$10.5 and $\sim$13.5 mag, with a peak at $\sim$12.5 mag. About half of them are impacted by an E(B-V) interstellar reddening lower than 0.5 mag, while for the other half, whose galactic latitude is smaller, the reddening extends up to 1.0 mag.

Most of the new dLHdC stars are located in the crowded area around the Galactic bulge where the mean GAIA eDR3 parallax uncertainty for sources with five-parameter solutions increases notably, mainly because only a few quasars are identified at small Galactic latitudes \citep[Fig.2]{2021A&A...649A...4L}. The parallax S/N measured is lower than 10 for 25 out of the 27 new dLHdC stars; the exceptions are A183 and A811, which are detected with a parallax S/N of $\sim$25.7 and $\sim$10.5 (Table~\ref{tab.newHdC}), respectively. The parallax S/N even decreases to values lower than 5 for 10 of them, which is nevertheless in the expected range at the distance of the Galactic bulge if we consider the average GAIA eDR3 parallax uncertainties of 0.02-0.03 mas for sources brighter than G=15 mag.

For distance estimates, we used the geometric distances inferred by \citet{2021AJ....161..147B} for the 1.47 billion GAIA eDR3 stars published with parallaxes. We list these distances as well as their 1$\sigma$ associated error in Table~\ref{tab.newHdC}. Nineteen of the new dLHdC stars are located at a distance ranging between 6 and 11 kpc, which again supports their Galactic bulge association. It is worth noting that two of the previously known dLHdC stars, HD 182040 and HD 137613, are the closest of all known HdC stars with distances of $\sim$884 pc and 1.2 kpc, respectively, closer than the three closest RCB stars, \object{R CrB}, \object{XX Cam}, and \object{RY Sgr}, whose respective distances are $\sim$1.3, $\sim$1.3, and $\sim$1.6 kpc.

We obtained the absolute V magnitudes and the unreddened (V-I)$_0$ colours for all dLHdC stars, the known EHe stars, some of the Magellanic RCB stars, and 31 of the known Galactic RCB stars that have an eDR3 GAIA measured parallax with an S/N higher than 2 $\sigma$ (only 3 of them have a parallax S/N between 2 and 3). We kept only the RCB stars for which we were able to convincingly estimate their maximum brightness in both V and I photometric bands. We used the light curves of the OGLE surveys for the Magellanic RCBs \citep{2008AcA....58..187U} and mainly the ASAS-3 and ASAS-SN surveys for the Galactic RCB stars, in association with the AAVSO database for the bright ones. For the dLHdC and EHe stars, we used the ATLAS All-Sky Stellar Reference Catalog \citep{2018ApJ...867..105T} to recalculate V and I magnitudes, and we used the ASAS-3 and ASAS-SN light curves to confirm the V magnitudes for all dLHdC stars. For the four really bright known dLHdC stars, we mostly used the magnitudes provided either by the UBVRIJKLMNH Photoelectric Catalogue \citep{1978A&AS...34..477M} or the catalogue of Stellar Photometry in Johnson's 11-colour system \citep{2002yCat.2237....0D}. To correct for interstellar reddening, we primarily used the 3D dust map produced by \citet{2019ApJ...887...93G}, but then used the E(B-V) reddening values produced by \citet{2011ApJ...737..103S} over the entire sky if information was not available in Green et al.

The result is presented in the M$_V$ versus (V-I)$_0$ colour-magnitude diagrams shown in Figure~\ref{fig_MV_VI}. The EHe stars, which are thought to be the final phase of HdC stars, are indeed bluer than any HdC stars and become even bluer while decreasing in magnitude. Surprisingly, we found that the known dLHdC stars and the majority of the new dLHdC stars are intrinsically fainter than not only the Magellanic RCB stars, but also most of the Galactic RCB stars by about 1.5 mag. Originally, \citet{2001ApJ...554..298A} and \citet{2009A&A...501..985T} brought to light a relation between the maximum brightness of the Magellanic RCB stars and their respective visual colour, but no such large spread was then observed in brightness. The outliers that are fainter RCB stars that did not follow the relation in \citet[Fig.3]{2009A&A...501..985T} were then explained by a need for a supplementary carbon extinction as their maximum magnitudes were not observed. Here, we accumulated enough measurements in the light curve of each RCB star used in the diagrams to be convinced of our measures of their respective maximum magnitudes. Overall, we found that the absolute brightness of HdC stars spans over $\sim$3 magnitudes for a colour (V-I)$_0\sim0.5$ mag.

\begin{figure}
\centering
\includegraphics[width=3.5in]{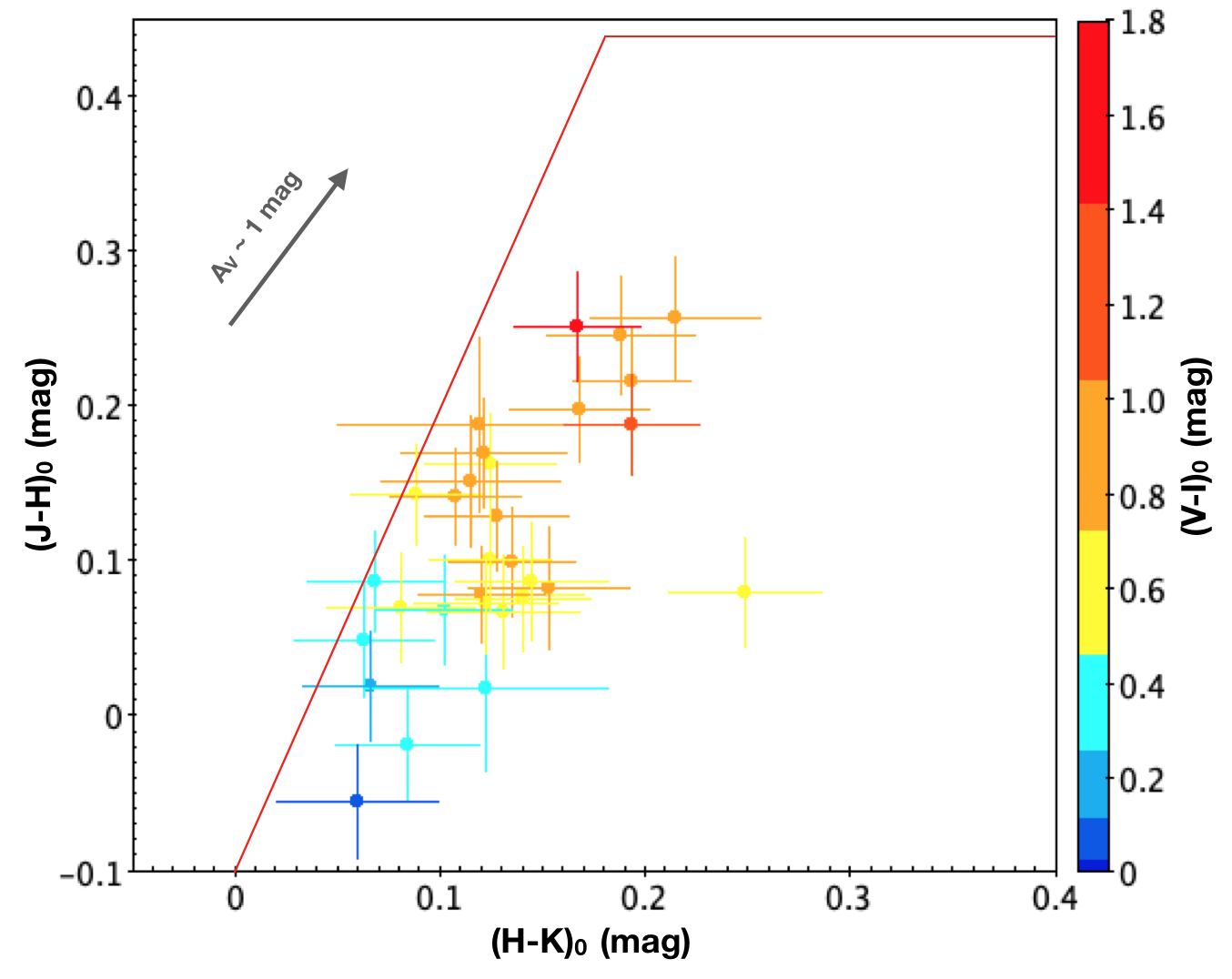}
\caption{Colour-colour J-H vs H-K diagram with 2MASS magnitudes corrected for extinction showing the positions of all dLHdC stars, similar to Figure~\ref{fig_Select-2MASS}. Here the dots and error bars are colour-coded with the respective (V-I)$_0$ colour found for each star. The applied selection cuts are represented with solid red lines.}
\label{fig_JHvsHK_V-I0}
\end{figure}

The  colour (V-I)$_0$ index of most of the dLHdC stars is lower than 1.0 mag, indicating temperatures between 5000 and 8000 K. We looked for reasons why we might have missed their cooler counterparts, but found no strong arguments for a selection effect. First, these colder objects are expected to be intrinsically fainter than the warmer ones, but from the second cut we applied (Fig.~\ref{fig_Select-GAIA}, right), the selected range in colour and brightness should be wide enough to be able to detect some of this missing population of cold dLHdC stars even if they were 1 mag fainter than their warmer counterparts. This is supported by our discovery of A166, the coldest new HdC of our sample (see the discussion of this star at Section~\ref{sec_A166}). Second, we looked at the effect of our first pragmatic selection cuts applied in the J-H versus H-K diagram (Eq.~\ref{eq.cut1}), and we realised afterwards that the distribution of the newly discovered dLHdC stars in this diagram depends on their effective temperature, as shown in Figure~\ref{fig_JHvsHK_V-I0}. There is no clear evidence that cool dLHdC star candidates might have been removed at this stage.

\begin{figure}
\centering
\includegraphics[width=3.5in]{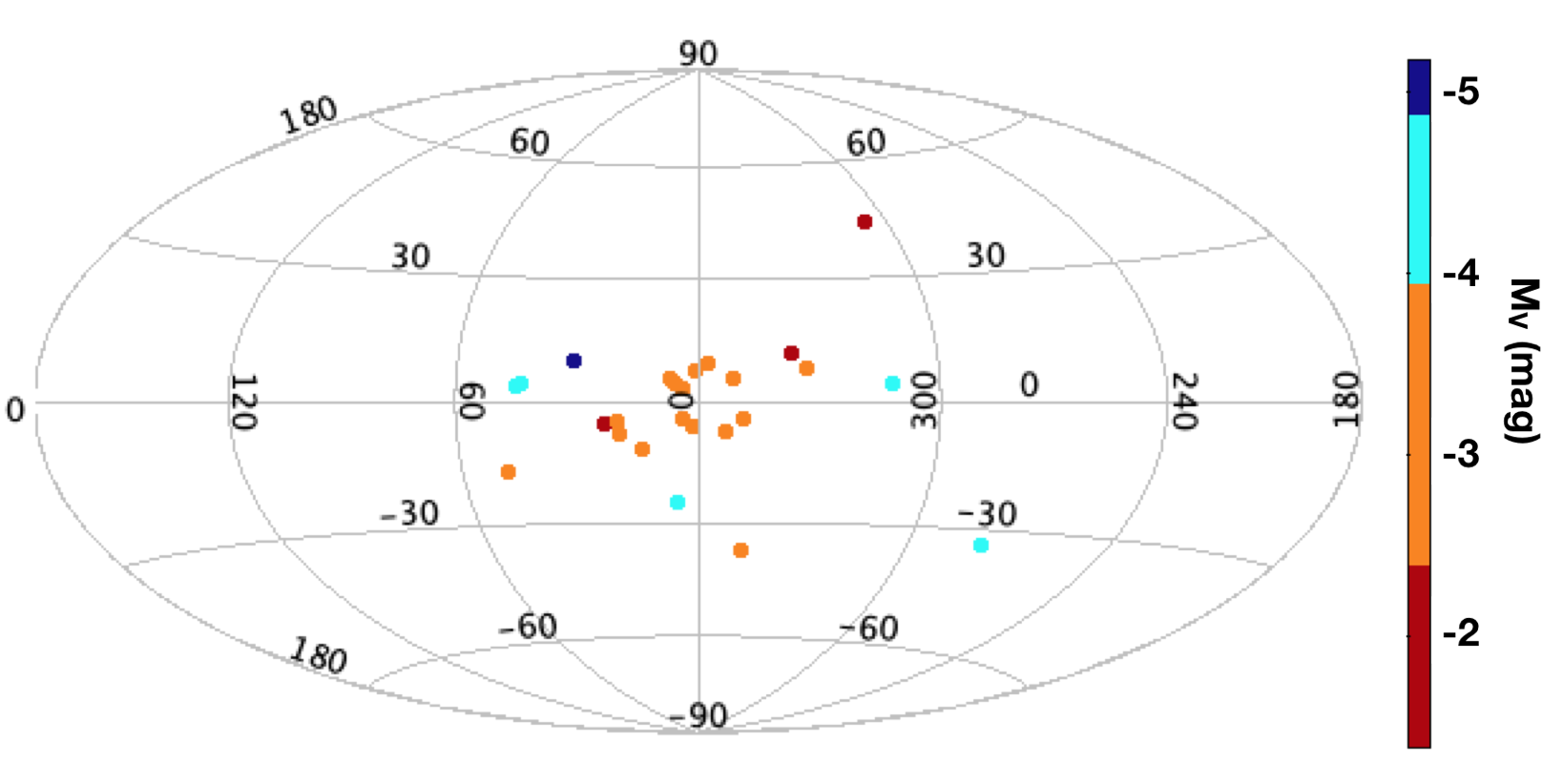}
\caption{Spatial distribution in Galactic coordinates of the 27 discovered dLHdC stars, colour-coded with their respective absolute magnitude M$_V$. A166 and the bright dLHdC stars are not located towards the sky area of the Galactic bulge, where we found most of the population of fainter dLHdC stars (M$_V\sim$-3 mag).}
\label{fig_GalacticDistrib_Mv}
\end{figure}

We found that the absolute magnitude of many dLHdC stars is about M$_V\sim$-3 mag, and they are distributed in a region of the HR diagram without known RCB stars. Furthermore, the six brightest dLHdC stars (M$_V<$-4 mag), that is, C526, F152, F75, A980, A814, and C542, are not distributed in a sky area towards the Galactic bulge like most of the other fainter dLHdC stars (see Fig.~\ref{fig_GalacticDistrib_Mv}). These fainter dLHdC stars could thus be an underlying population of real long-term dustless HdC stars.

On the warmer side, we found that three hot RCB stars, \object{MACHO 11.8632.2507}, \object{DY Cen}, and \object{MV Sgr}, are located in the same part of the diagram, with similar absolute magnitudes M$_V\sim$-3 mag and colours (V-I)$_0\sim$0.15 mag, as expected from their effective temperatures, which are higher than 12000 K. The colour indices of four warm HdC stars were lower than or similar to those of these hot RCB stars. They are the two RCB stars, \object{Y Mus} and \object{[TCW2013] ASAS-RCB-10}, and the two new dLHdC stars, B564 and A980. We almost certainly applied an interstellar reddening correction that is too high for these objects, whose effective temperature is about 8000 K. We indicate in the colour-magnitude diagram the over-correction needed to reach a more reasonable colour of (V-I)$_0\sim$0.25 mag.

Similarly, on top of the 1$\sigma$ error bars that represent our best knowledge of the uncertainties related to distance, apparent maximum brightness, and colours, the effects of an over- and/or under- correction of interstellar dust reddening in the M$_V$ versus (V-I)$_0$ colour magnitude needs to be imagined. Extinction vectors are indicated in the two diagrams of Figure~\ref{fig_MV_VI}. We do not expect a drastic change in the overall distribution of HdC stars. However, we noted that for few of them, a shift up to 0.2 mag in colour is expected when their (V-I)$_0$ colour indices are compared with their temperature estimates based on their respective spectra. We found, for example, that C17 (classified as a mild dLHdC star) has an estimated (V-I)$_0$ colour index of 0.47 mag that is too blue for its temperature. Conversely, A182 (classified as a warm dLHdC star) has an estimated (V-I)$_0$ colour index of 0.63 mag that is too red, and a higher reddening correction should thus be applied. Finally, the warm dLHdC star, A182, should probably be bluer than its derived colour of (V-I)$_0\sim$0.63 mag.

The GAIA eDR3 parallax S/N of the two new HdC stars F152 and F75  is lower than 3, but their geometric distance estimate associated with a larger 1$\sigma$ error is still statistically valid. We plot these two in the colour-magnitude diagram of Fig.~\ref{fig_MV_VI} to illustrate their positions compared to the other dLHdC stars as we have evidence of dust production activity for these two stars despite the observational lack of a warm circumstellar dust shell around them (see Section~\ref{sec_dust}). They both present a spectrum that indicates a warm effective temperature of $\sim$7500 K, and their positions in the HR diagram correspond to those of RCB stars of the same temperature with absolute M$_V$ magnitudes ranging between -4.4 and -5 mag for a (V-I)$_0$ colour index of $\sim$0.4 mag.

Also, in the curiosity cabinet, we note that A980 and B564 are the two warmest dLHdC stars we found, with a temperature of $\sim$8000 K. They present very similar spectra (Fig.~\ref{fig_Spectra_newHdC_warm}), but their absolute M$_V$ magnitude departs by 1.5 mag, A980 being the brighter. We were unable to distinguish any differences in the absorption line equivalent widths within our spectroscopic resolution.
 
\subsection{Interpretation in the framework of the double-degenerate scenario}

In the double-degenerate scenario, \citet[Fig.2]{2002MNRAS.333..121S} have presented the evolution of an HdC star in the HR diagram following the merger of a CO-WD with an He-WD for different initial total masses and WD mass ratio. They showed that such an object becomes brighter and warmer after an initial cool phase following the merger and that the maximum luminosity is expected to increase with initial total mass. They calculated a difference in maximum luminosity of $\sim$2.1 mag between a WD system with a total mass of 0.6 M$_{\sun}$ and one of 0.9 M$_{\sun}$. In light of these models, we might here see for the first time the evolutionary sequences of WD-binary mergers with a wide range of initial total masses.

\begin{figure}
\centering
\includegraphics[width=3.5in]{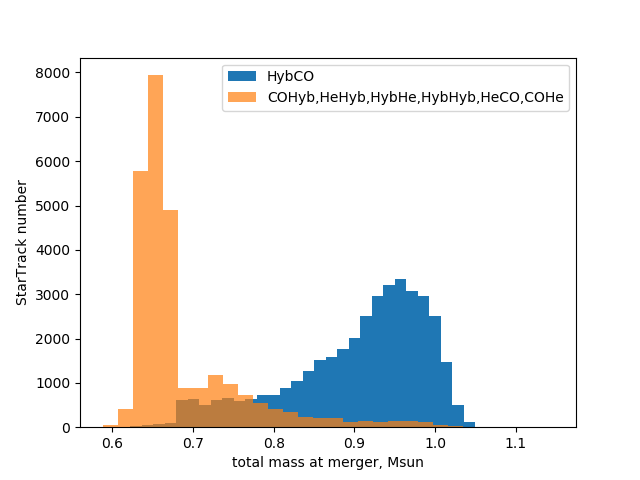}
\caption{Plausible mass distribution of the total WD systems that are considered to be the progenitor of HdC stars. The contributions from different WD merger types are considered in all combinations between a hybrid, a CO, and a He WD (except for CO+CO and He+He). For illustration purpose, we separate the more massive channel made of a hybrid and a CO WD merger (blue) from all other ones (orange) listed in the legend. The overall distribution is bimodal.}
\label{fig_WDmassdistrib}
\end{figure}

The mass distribution is predicted by population synthesis simulations of close binary systems in \textit{StarTrack} \citep{2008ApJS..174..223B,2014MNRAS.440L.101R} after various mass transfer phases. The double WD merger types with a high total mass (CO-CO, CO-ONe, or ONe-CO) and those with a low total mass (He-He) are neglected as a reasonable channel to be HdC progenitor stars. Only systems with intermediate total WD masses are used here. In these double WD binaries, a hybrid COHe WD \citep{1996MNRAS.280.1035T,2019MNRAS.482.1135Z} forms before a CO WD contributes to the massive end of the total mass distribution, with an average merger mass of $\sim$0.9 M$_{\sun}$ \citep[Fig.1]{2015ApJ...809..184K}. These types of WD mergers are also more numerous compared to the other WD merger types we consider and are formed through a stable Roche-lobe overflow (RLOF) phase when the primary star fills its Roche lobe on the MS or in the Hertzsprung gap, continuing through to when the star is a red giant, followed later by a common-envelope phase where the secondary star loses its envelope as a red giant or an AGB star. Some systems undergo a further mass transfer phase, typically dynamically stable, where the donor (having previously lost its hydrogen envelope in RLOF) is on the helium main sequence. The merger mass distribution from StarTrack is presented in Figure~\ref{fig_WDmassdistrib}. The distribution covers a total mass range between $\sim$0.6 and $\sim$1.05 M$_{\sun}$ and has a bimodal structure. The notation HybCO indicates that the initially more massive star on the zero-age main sequence (ZAMS) formed the hybrid WD, where COHyb indicates that the initially more massive star on the ZAMS formed the CO WD. For clarity, we show the less-populated WD merger channels grouped together (in orange). Mergers between a helium WD and a hybrid WD produce average merger masses on the lower end, $\sim$0.65 M$_{\sun}$, thus may contribute more to the less luminous population of HdC stars. Here, a low metallicity that is more along the lines of what is found in the SMC was used, although the main distribution features persist at solar metallicity. 

Although likely not all of the WD merger channels we explored contribute to the HdC star population (e.g. some might contribute to low-luminosity thermonuclear supernovae, as explained in \citet{2017NatAs...1E.135C} or \citet{2021MNRAS.503.4734P}), the different formation scenarios and resulting WD merger masses make it rather tempting to assign different formation pathways to the different HdC star populations.

We also found that WD mergers from the more massive HybCO channel also have an interesting bimodal delay time distribution with about 40\% of systems merging within 2 Gyr after star formation (with the first mergers occurring 250 Myr after starburst), while about 60\% take at least 5 Gyr after star formation to merge with some systems capable of merging beyond a Hubble time. Those with shorter delay times encounter a common-envelope phase when the mass-losing star is a red giant, whereas those with longer ($>$5 Gyr) delay times encounter the common-envelope phase when the mass-losing star is on the AGB. In the hypothesis that all WD mergers from this channel form HdC stars, we thus would expect to find those to be present both in the Galactic bulge as well as regions of active star formation. We do not have strong evidence of a large population of HdC stars located in the spiral arms, although their progenitors may still be actively forming there. Using the 3D dust map of Green et al. (2019), we noted that the three following RCB stars, GU Sgr, UV Cas, and V2552 Oph, and the dLHdC star A223 have a Galactic position in which we found a large increase of dust reddening consistent with a location within or near a spiral arm. 


As expected in the double-degenerate scenario, it is also interesting that we can interpret the upper luminosity envelope in the lower panel of Figure~\ref{fig_MV_VI}, formed by the maximum absolute magnitudes recorded for each bin of (V-I)$_0$ colours, starting at (M$_V$, (V-I)$_0$)$\sim$(-5.2 ; 0.3) mag, and ending at $\sim$(-3.3 ; 2.0) mag, as the evolutionary sequence of HdC stars with the highest total initial mass possible to form such supergiant stars. This envelope is illustrated as a dotted curve in Figure~\ref{fig_MV_VI} and corresponds to the simple second-order polynomial function M$_V$=A+B$\times$(V-I)$_0$+C$\times$(V-I)$^2_0$ with A=-5.08, B=-0.539, and C=0.711. Above this mass threshold, the WD mergers may result in supernovae \citep{2010ApJ...714L..52S,2011MNRAS.417..408R,2014MNRAS.440L.101R}. Curiously, however, for cool temperatures (i.e. (V-I)$_0>$1.0 mag), this envelope consists of Magellanic RCB stars alone, all cold Galactic RCB stars are fainter. We cannot explain these differences by any issues due to the calibration of the photometric V and I bands, the interstellar reddening correction (most cool Galactic RCB stars are located towards low-reddening sky areas), or the distance estimates. It is based on 7 cool Galactic RCB stars for which we succeeded in gathering enough information. More work will be needed on the more than 60 other known cool Galactic RCB stars to confirm this observation. This contrast might be interpreted in terms of a difference of stellar populations between the two galaxies, resulting in two similar total initial mass regimes, but two distinct WD mass ratios. This scenario is presented by \citet[Fig.2]{2002MNRAS.333..121S} in the specific case of a total initial mass of 0.9 M$_\sun$. The evolutionary track resulting from a 0.6 M$_\sun$ CO-WD that accreted a 0.3 M$_\sun$ He-WD is brighter at cool temperatures than the one resulting from a 0.5 M$_\sun$ mass CO-WD that accreted a 0.4 M$_\sun$, but both tracks reach similar luminosities at warmer temperatures before entering the EHe phase. Furthermore, a theoretical study of the metallicity dependence of the initial to final mass relation made by \citet{2015MNRAS.450.3708R} shows that metal-rich progenitors result in less massive WD remnants because the mass-loss rates increase when associated with high metallicity values. Differences up to $\sim$0.1 M$_\sun$ are expected in the final WD mass. This effect may thus explain our observations, because the Magellanic stellar population has a lower metallicity than the Galactic one, it would result in more massive Magellanic RCB stars than their Galactic counterparts.

\subsection{Photometric variabilities \label{sec_variability}}

We studied the light curves of all dLHdC stars aggregated from the following four monitoring surveys: ASAS-3 \citep{1997AcA....47..467P}, ASAS-SN \citep{2014ApJ...788...48S,2017PASP..129j4502K}, Catalina \citep{2012IAUS..285..306D}, and Palomar Gattini IR\footnote{The Gattini IR light curves are available at this URL: \url{http://rcb.iap.fr/tracking_dLHdC/Gattini-IR/}} \citep{2020PASP..132b5001D,2019NatAs...3..109M}. The ASAS-SN light curves were used to study variabilities on short timescales, while the others were inspected to check for the presence of possible declines.

The striking first result from this study is that most of the dLHdC stars (19 out of the 32) show no variability at the photometric resolution of the ASAS-SN survey, while for the other 13, we detect only some low peak-to-peak amplitude variabilities (between 0.05 to 0.15 mag) on a timescale of about 10 days (see Table~\ref{tab.VariabilityHdC}). This is different from the irregular oscillations observed in RCB stars, which vary when they are at maximum luminosity between 0.2 and 0.4 mag on a timescale between 20 to 60 days. This indicates clear physical differences between these two populations of HdC stars.

The low-amplitude photometric variability of the dLHdC stars is supported by other surveys. We note that using the Hipparcos dataset, \citet{2012MNRAS.427.2917R} found that HD 137613 presents variabilities with a peak-to-peak amplitude of 0.05 mag on a timescale of 0.1 day, but this is not confirmed by the KELT survey \citep{2018AJ....155...39O}, which did not find variabilities greater than 0.04 mag for that star on timescales ranging from 30 mins to one day. \citet{2012MNRAS.427.2917R} classified the brightest dLHdC star, HD 182040, as a variable star with a peak-to-peak amplitude of 0.04 mag and a period of $\sim$1.3 days. This variability is confirmed by the Mascara monitoring survey of bright stars \citep{2018A&A...617A..32B}, which reports a clear variability signature that is irregular, but with some oscillations observed with 0.04 mag peak-to-peak amplitude and a duration of $\sim$9 days. The new dLHdC stars A183, C20, A980, and A166 are listed by \citet{2018AJ....155...39O} as non-variable stars on a short timescale of less than one day with amplitudes that would be higher than $\sim$0.01, $\sim$0.03, $\sim$0.025, and $\sim$0.05 mag, respectively. However, on a longer timescale, the ATLAS survey \citep{2019yCat..51560241H} reports that A166 might be an irregular variable with $\sim$0.1 mag variation, consistent with our observations from its ASAS-SN light curve (Table~\ref{tab.VariabilityHdC}). The ATLAS\ survey also reports that A223, B566, C542, and A977 show long-timescale variations of between 0.07 and 0.1 mag amplitude, without being successful in a clear classification of variability class. Finally, the TASS Mark IV photometric survey of the northern sky \citep{2006PASP..118.1666D} lists A980 with a variability index indicating significant change in brightness, and B42 is listed in the ASAS-SN catalog of variable stars \citep{2018MNRAS.477.3145J} to present variabilities of $\sim$0.09 mag amplitude on a timescale of $\sim$41 days. We also confirm these observations. For B42, we also observe multiple irregular variations on a timescale of $\sim$10 days (Table~\ref{tab.VariabilityHdC}).

\begin{figure}
\centering
\includegraphics[width=3.5in]{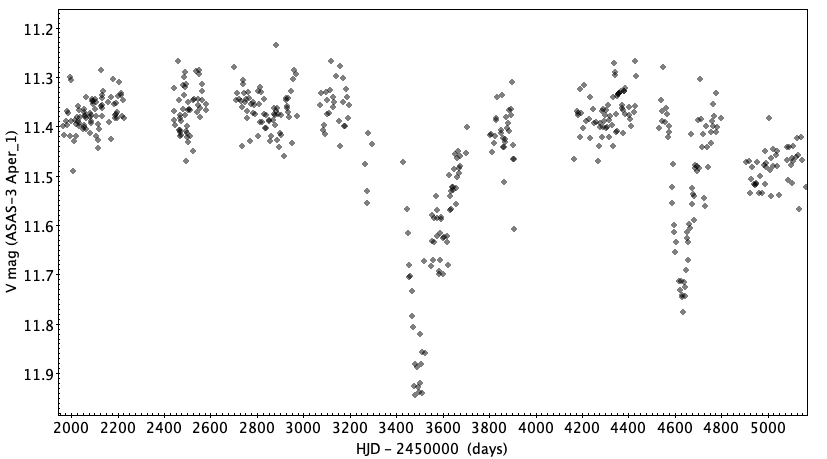}
\caption{ASAS-3 V band light curve of F75. Two small photometric declines were observed.}
\label{fig_lc_F75}
\end{figure}

\begin{table}[]
\caption{Photometric variabilities observed in the ASAS-SN light curves of dLHdC stars and RCB stars.
\label{tab.VariabilityHdC}}
\medskip
\centering
\begin{tabular}{ccc}
\hline
Star  &  Timescale &  Peak-to-peak\\
  &   (days)  &  amplitude (mag) \\
\hline
\multicolumn{3}{c}{\emph{dLHdC stars}}\\
\hline
A166  & $\sim$25 \& $\sim$200  &  0.15  \\ 
A182 & $\sim$15  & 0.15  \\ 
A223  &  $\sim$10  &  0.1 \\
A249 & 10-20  & 0.05  \\
A770 & $\sim$10   &  0.1 \\
A980 & 10-20  & 0.1   \\  
B42 & 10-40 &  0.1  \\ 
B564 & $\sim$10  &  0.08  \\
B566 & $\sim$10  &  0.12 \\   
C27 & 10-40   & 0.15  \\ 
C526  & 10-40   & 0.15   \\ 
F75 &  10-25  &  0.15 \\
F152 & 30-60  & 0.1-0.2  \\
\hline
\multicolumn{3}{c}{\emph{RCB stars}}\\
\hline
C105 & 40  & 0.4  \\ 
HD 175893 & 20  &  0.3  \\
\hline
\end{tabular}
\end{table}

The ASAS-SN light curve of the new RCB star, C105, shows variabilities typical of RCB stars with a peak-to-peak amplitude of $\sim$0.4 mag and a timescale of 40 days. It was listed in the ASAS-SN catalogue of variable stars as a non-periodic variable star with an amplitude of $\sim$0.26 mag \citep{2018MNRAS.477.3145J}. This is also the only object discovered in the present analysis that was listed as a variable star in the SIMBAD database due to the high-frequency light curve published by the HATNET variability survey \citep{2004AJ....128.1761H}, which shows a smooth variation of 0.12 mag amplitude over 30 days. Finally, \citet{2019yCat..51560241H} reported variations of $\sim$0.35 mag detected by the ATLAS survey. We also studied the ASAS-SN light curve of HD 175893 and found variabilities of 0.3 mag on a timescale of 20 days (Table~\ref{tab.VariabilityHdC}).

We did not detect photometric declines in any of the dLHdC stars except one. The ASAS-3 light curve of F75 presents two declines at JD$\sim$2453450 and $\sim$2454600 days that last $\sim$300 and $\sim$150 days, respectively, with a maximum obscuration of $\sim$0.55 mag in the first event and $\sim$0.40 mag in the second (see Fig.~\ref{fig_lc_F75}). The recovery part of the first decline and the second decline were also observed by the Catalina survey. These declines certainly indicate an ongoing phase of dust production. This is discussed below when we study F75 mid-IR WISE photometry. F75 may be in the process of starting its phase as an RCB star.

\subsection{Dust around the new dLHdC stars? \label{sec_dust}}

The RCB stars are the dusty version of HdC stars. They are known to be actively producing dust made of amorphous carbon dust grains and thus are surrounded by warm circumstellar shells \citep{1997MNRAS.285..317F,2011ApJ...739...37G,2012A&A...539A..51T}. The RCB stars produce dust at all temperatures, from very cold (T$_{eff}\sim$4000 K) to very hot (T$_{eff}>$12000 K). The mechanisms needed to make this dust therefore appear not to be much influenced by temperature. Moreover, a few RCB stars are surrounded by multiple shells. We observed these features in the spectral energy distribution (SED), as seen by WISE in 2010, of the three following hot RCB stars: \object{MV Sgr}, \object{DY Cen}, and \object{MACHO 11.8632.2507}, but also of a cold and of a warm RCB star, \object{WX CrA} and \object{UW Cen}, respectively \citep{2012A&A...539A..51T}. This shows that RCB stars of any temperature can stop or slow their production of dust for a period of time before restarting it again. Finally, \citet{2018AJ....156..148M} presented high-resolution mid-IR images of a few RCB stars showing the existence of extended cold shells. The dust production may be active throughout the lifetime of an RCB star.

The four historic dLHdC stars are known to be deprived of any warm circumstellar inner dust shell, unlike RCB stars since the data release of the WISE mid-IR photometry and thus since the study of their SED up to 22 $\mu$m \citep{2012A&A...539A..51T}. Before this, we simply knew that these four HdC stars were never detected to undergo large photometric declines. Consequently, they may be thought to either not have started a dust production process yet or perhaps this process stopped long ago. At longer wavelength, searches were made to detect the possible remains of cool extended dust shells. This is made difficult by the background dust emission, which is often dominant at low Galactic latitudes. First, using the IRAS satellite catalogue, \citet{1985A&A...152...58W} searched for IRAS flux, but HD 182040 was only detected at 12 $\mu$m. Then, recently,  \citet{2018AJ....156..148M} studied the dLHdC star HD 173409 using the Herschel/PACS and SPIRE instruments, and found that no nebulosity is visible in the images taken between 70 and 500 $\mu$m. Around most RCB stars they observed, however, an extended cold dust shell was discovered in addition to the existing warm inner shell. This is another clear difference between the two populations of HdC stars.

We examined the mid-IR WISE colours of each newly discovered HdC star to detect the possible presence of circumstellar shells. We found that, in contrast to all known Galactic RCB stars located in similar crowded areas, no detection is reported for most of the new HdC stars in the reddest of the four WISE bands: indeed 18 out of the 27 dLHdC stars have a magnitude limit in the [22] photometric band. Additionally, no entry in the WISE All-Sky and ALLWISE catalogues was found for B566. The nearest WISE source is located at 2.2 arcsec from B566 and corresponds in fact to the blend of B566 with a nearby object. Finally, the faintest known dLHdC star, HE 1015-2050, was reported with two magnitude limits in the two redder WISE bands. Neither object is part of our study. In addition to the four brightest known dLHdC stars, 8 of the newly discovered stars have a valid measurement in all four WISE passbands.

\begin{figure}
\centering
\includegraphics[width=3.5in]{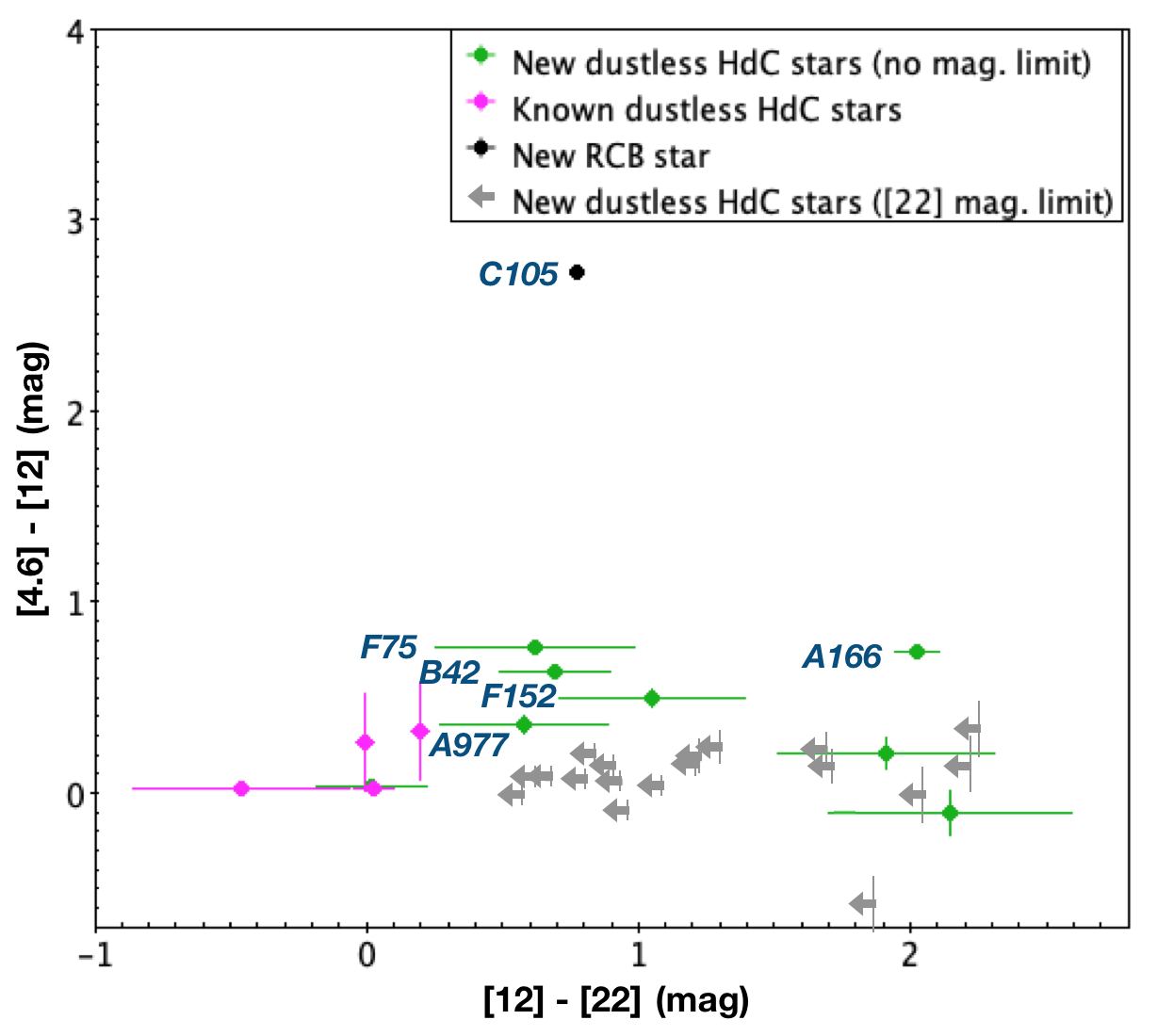}
\caption{Colour–colour $[4.6]-[12]$ vs $[12]-[22]$ diagram using WISE ALLWISE mid-IR magnitudes; see \citet[Fig.1]{2020A&A...635A..14T} for a direct comparison with known RCB stars. The new RCB star, C105, is indicated with a black dot. The new dLHdC stars are indicated with green dots and grey arrows for those published with a magnitude limit in the $[22]$ passband. The known dLHdC stars are shown with purple dots.}
\label{fig_W2W3vsW3W4}
\end{figure}

\begin{figure}
\centering
\includegraphics[width=3.5in]{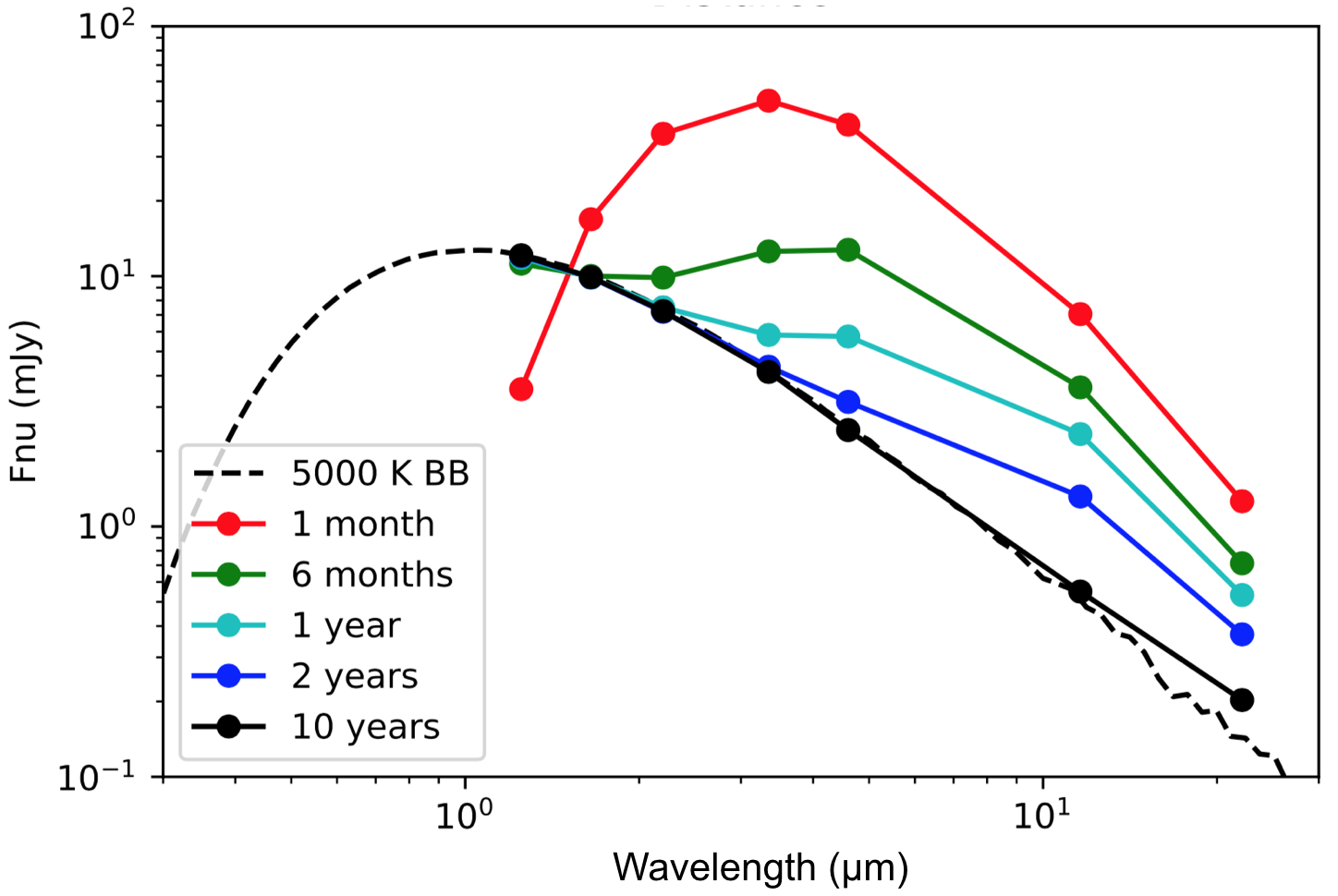}
\caption{Simulation of the overall SED of a 5000K RCB star with a known episode of strong dust production of 10$^{-8}$ M$_\sun$ at t=0. The models were made using the 3D Monte Carlo radiative transfer code Mocassin \citep{2005MNRAS.362.1038E} with a homogeneous distribution of amorphous carbon dust with the size distribution of \citet{1989ApJ...345..245C}. The stellar radius is 100 R$_\sun$. The seven dots that define the dust shell SED correspond to the 2MASS and WISE passbands. The J-band luminosity of the one-month-old dust shell model is lower than the luminosity of the photosphere blackbody due to the extinction effect. It corresponds to the characteristic RCB star photometric declines, here of 1.2 mag in the J band.}
\label{fig_RCB_SED}
\end{figure}

\begin{figure}
\centering
\includegraphics[width=3.3in]{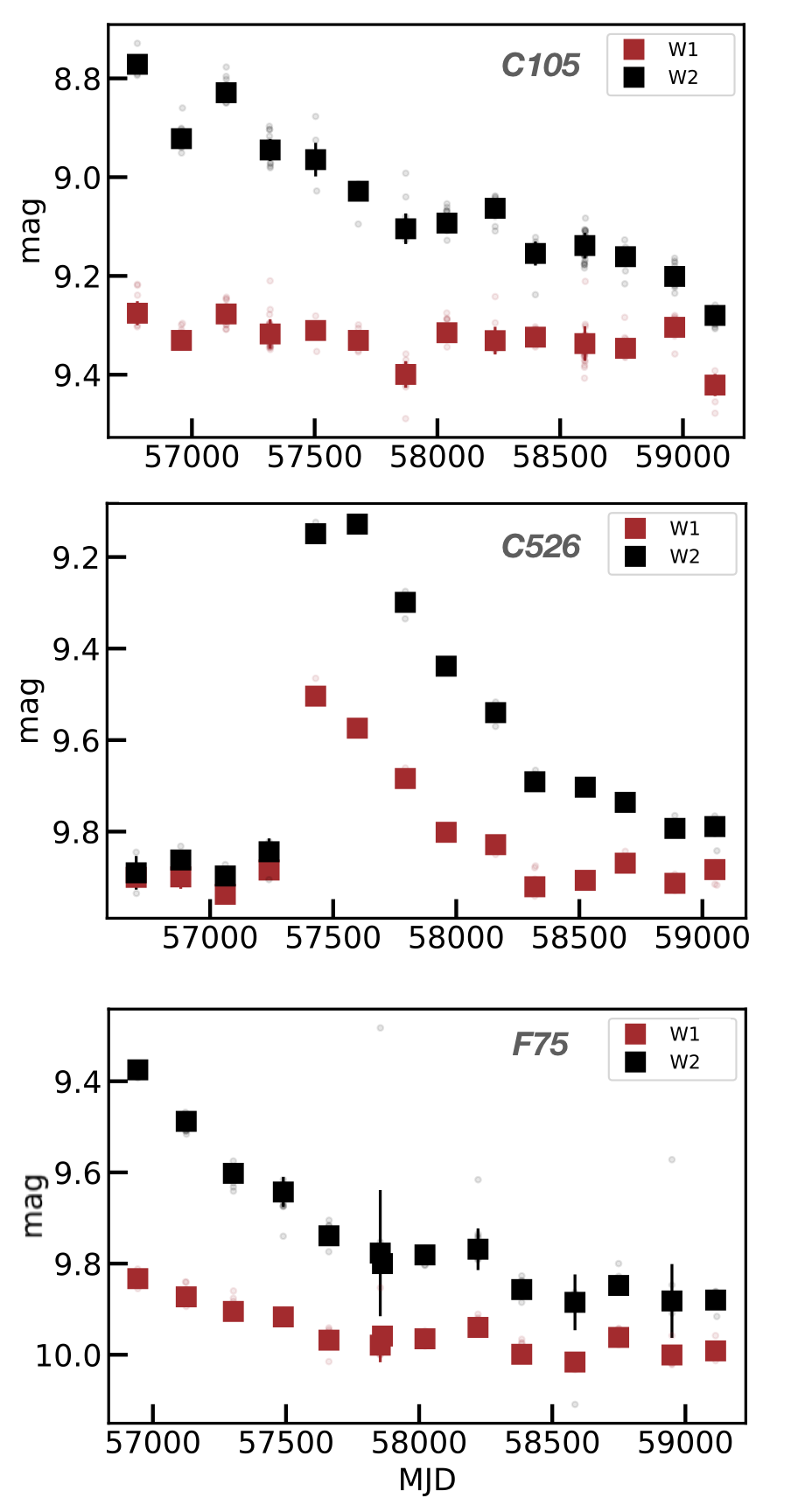}
\caption{Mid-IR light curves (2013-2021) from the NEOWISE survey of C105, C526, and F75. All three new warm dLHdC stars show mid-IR variabilities indicating a dust production activity from these stars. The sudden increase in fluxes observed in the W1 and W2 bands for C526 is the signal of rapid dust formation. The slow decrease rate of the W2 fluxes observed in the three stars is due to the cooling of the dust while it is dispersed in the circumstellar surrounding.}
\label{fig_Mid-IR_lc}
\end{figure}

Our study can be summarised with the WISE colour-colour [4.6]-[12] versus [12]-[22] diagram presented in Figure~\ref{fig_W2W3vsW3W4}. First, C105 immediately stands out as it is located in a part of the diagram away from the main locus of dLHdC stars. This is a new RCB star. Its position corresponds to the area expected for a typical RCB star that has formed a thick circumstellar dust shell with a temperature of $\sim$500 K (see \citet[Fig.1]{2020A&A...635A..14T} for a direct comparison with known RCB stars). Second, the mid-IR colours of all but five dLHdC stars are consistent with no IR excess. These five HdC stars are A977, F75, F152, B42 and A166. Their [4.2]-[12] colour is higher than 0.3 mag, F75 has the reddest colour with $\sim$0.76 mag. These excesses, if confirmed, should indicate the presence of thin dust shells.

We mentioned in the preceding section that the warm HdC star F75 has undergone two small photometric declines indicating a current phase of dust production. This is supported by its WISE colours. This suggests that F75 is certainly an HdC star that is in a transition time as it is starting its RCB phase. Interestingly, this is the first known HdC star of this type. The WISE colour of three of the four other HdC stars (A977, F152, and B42) is similar to that of F75, but slightly bluer, indicating they may only have started a similar transition. The small H$_\alpha$ emission line observed in the spectrum of F152 may be explained by the association of this possible new phase with a strong stellar wind. The variability observed in the light curve of F152 also corresponds to the typical variabilities observed in RCB stars at maximum brightness. For the cool HdC star A166, the situation appears quite different. Its $[12]-[22]$ colour index is redder by 1.5 mag. This implies the presence of a cold T$_{shell}\sim$200 K dust shell, which would argue in favour of a scenario in which the dust production has stopped for a while, and the dust is cooling whilst moving away from the star.

All the discussions above focus on the mid-IR colour, which mostly corresponds to one particular epoch, that is, the nine-month period of the WISE survey in 2010. As has been demonstrated by \citet{1997MNRAS.285..317F}, however, the mid-IR fluxes of RCB stars vary in time in relation to the sudden changes in dust production from the star. We present in Figure~\ref{fig_RCB_SED} the overall SED of a 5000K RCB star that was observed at different epochs (between one month to 10 years) following a large dust production of 10$^{-8}$ solar masses. We used MOCASSIN \citep{2005MNRAS.362.1038E}, a 3D Monte Carlo radiative transfer code, to make these models. The effect of the resulting dust shell moving away from the star at 400 km.s${-1}$ is visible as it cools and is dispersed into the interstellar environment. If an RCB star remains highly productive, a warm circumstellar dust shell is continually observed around it. Otherwise, almost no dust emission would be detected only 10 years after the production of such a high quantity of dust. We studied the light curves of all the dLHdC stars produced by the on-going mid-IR NEOWISE survey \citep{2011ApJ...731...53M,2014ApJ...792...30M}, which has been observing the entire sky in the two bluest WISE passbands since 2013. We found no variations during the 8 years of observation, and even similar W1 and W2 magnitudes were reported by both the WISE and the NEOWISE surveys for all but two dLHdC stars, F75 and C526, and for C105, our new RCB star. Their respective mid-IR light curves are presented in Figure~\ref{fig_Mid-IR_lc}. They all show a decrease in W2 fluxes with a total amplitude of $\sim$0.5, $\sim$0.6, and $\sim$0.5 mag for C105, C526, and F75, respectively, and a slighter decrease in the W1 passband of $\sim$0.1, $\sim$0.4, and $\sim$0.2 mag, respectively. This corresponds to the cooling of the dust that was recently newly produced near the star, as it disperses into the circumstellar environment. An emission due to a sudden dust production was even observed in 2016 for C526. In less than 200 days, the W2 flux increased by $\sim$0.8 mag, followed by a slow recovery to the value of the initial flux before the burst, which is also similar to the value reported by the 2010 WISE observations. The quantity of dust produced is here lower by a factor of $\sim$20, that is, $\sim$5$\times$10$^{-10}$ M$_\sun$, than in the model presented in Fig.~\ref{fig_RCB_SED}. For C105, the sudden dust emission occurred before 2010 as the maximum W2 (W1) flux observed in 2013 is $\sim$1.1 ($\sim$0.6) mag fainter than the flux observed in 2010. The total variation is thus $\sim$1.6 ($\sim$0.5) mag in W2 (W1). For F75, the dust emission occurred between the two surveys, as the reported WISE magnitudes were fainter by $\sim$0.4 ($\sim$0.3) mag in W2 (W1).

In conclusion, we have accumulated some evidence that shows that F75, C526, and to a lesser degree, F152, are in fact producing some dust at a low rate as it does not accumulate around these stars to form a long-lasting shell. These three warm stars have the same brightness (M$_V$ between -4.4 and -5.0 mag) as known RCB stars of the same temperature ($\sim$7500 K). They might even be typical RCB stars passing through a transition time, entering or leaving the RCB phase. We reached a similar conclusion for the only very cold HdC star we found, A166. A detailed discussion of this particular star is given in the following subsection.

\subsection{Specific case of A166 \label{sec_A166}}

The HdC star A166 has remained apart throughout our analysis of the other dLHdC stars. It has accumulated differences in almost all observational aspects. The differences start with its location on the sky as it is a halo star found to be at $\sim$6 kpc from the Sun. A166 is redder than all the newly discovered HdC stars revealed with a (V-I)$_0$ colour index of 1.8 mag, indicating an effective temperature of about $\sim$4000 K. It is located at a position in the colour-magnitude diagram near two known cold RCB stars (\object{[TCW2013] ASAS-RCB-4} and \object{V1157 Sgr}), away from the main locus of dLHdC stars (see Fig.~\ref{fig_MV_VI}). We note that its line of sight is only weakly impacted by interstellar dust reddening with E(B-V)$\sim$0.07 mag (similar to the two cold RCB stars).

Its cool spectrum (Fig.~\ref{fig_Spectra_newHdC_cool}) is also uncommon. It shows no CN features at all, but some C$_2$ band heads. Furthermore, along with being hydrogen deficient, it has the curious characteristic of strong absorption lines due to s-process elements such as strontium and barium (see a detailed discussion in \citealp{crawford_2022}).

\begin{figure}
\centering
\includegraphics[width=3.5in]{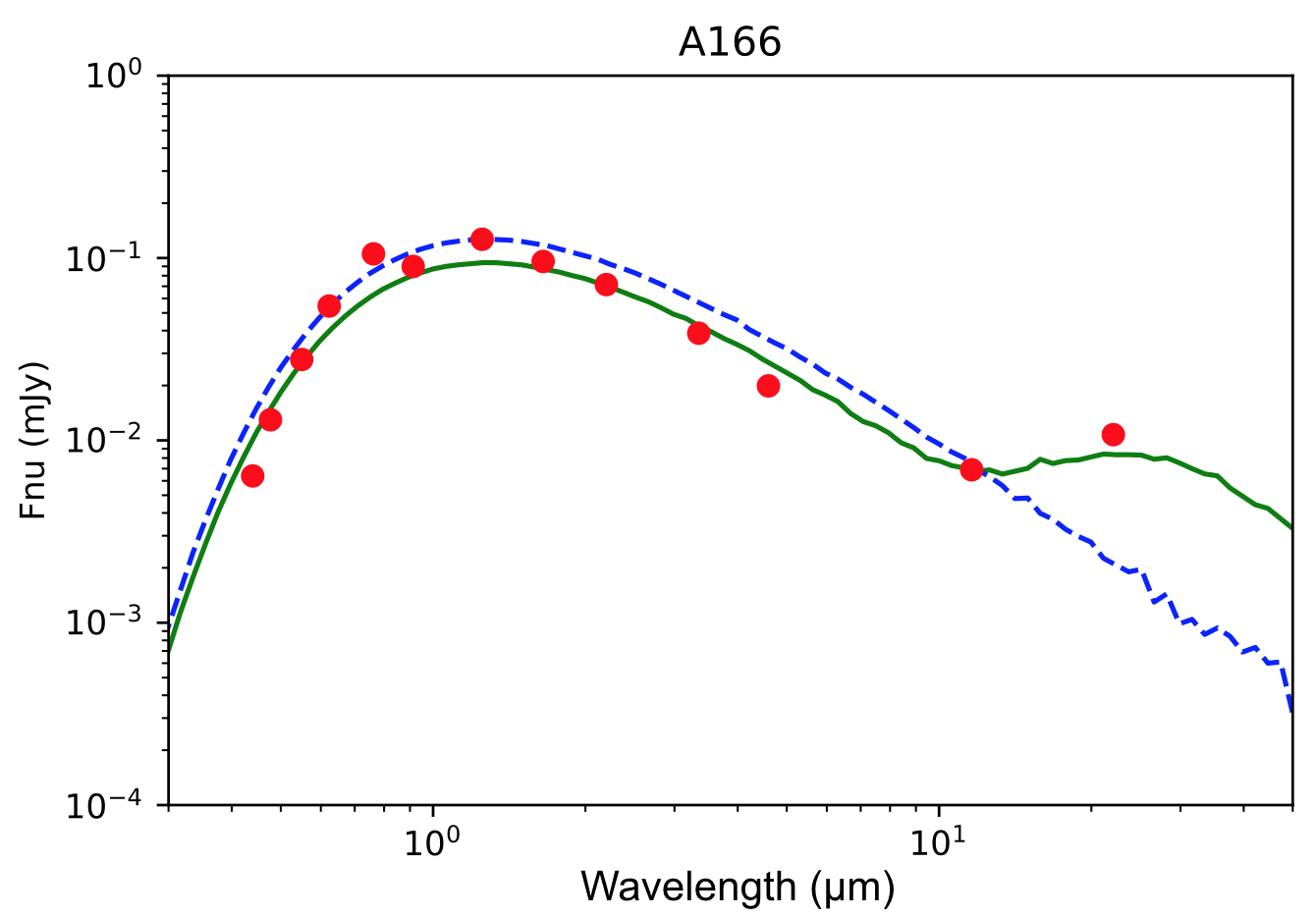}
\caption{Spectral energy distribution of A166. The red dots show the photometry. The dashed blue line shows a 4000 K blackbody curve. The green line adds 10$^{-8}$ solar masses of amorphous carbon dust with a typical \citet{1989ApJ...345..245C} size distribution at a distance of 6.4$\times$10$^{15}$ cm ($\sim$430 AU). It would take five years for the dust to reach this distance if its velocity were 400 km s$^{-1}$. }
\label{fig_A166_SED}
\end{figure}

Then, we studied its entire spectral energy distribution in detail, we found an IR excess in the two reddest WISE mid-IR photometric bands (Fig.~\ref{fig_W2W3vsW3W4}). This reveals the presence of a cold dust shell with a temperature between 200 and 300 K (Fig.~\ref{fig_A166_SED}) that is not seen in any other dLHdC stars (see Section~\ref{sec_dust}). The photometric variability observed in all light curves we accumulated shows irregular variabilities, but patterns with two long time-scales of $\sim$25 and $\sim$200 days are visible, and a peak-to-peak amplitude of $\sim$0.15 mag. This is not seen in any other dLHdC stars. Their observed variabilities are of smaller amplitude and do not last as long.

Finally, \citet{Karambelkar_2022} reported the possible detection of blueshifted He I lines in the IR spectrum of A166. This attribute is typically observed in RCB stars as it is due to a dust-driven wind \citep{2013AJ....146...23C}. Again, no other dLHdC star shows a similar pattern. Moreover, the \citet{Karambelkar_2022} measurement of the oxygen 16 to 18 isotopic ratio indicates that its enrichment in the oxygen 18 isotope is not as large as they observed for dLHdC stars, but corresponds more to ratios found in RCB stars.

In conclusion, A166 probably also is an HdC star that is in a transition time because it lacks the typical warm circumstellar dust shell, but has a cold shell that is moving away from the star. If we consider a dust velocity of 400 km s$^{-1}$, the production level needed to sustain a warm dust shell must have stopped about five years before the 2010 WISE survey. As no noticeable mid-IR luminosity variations are observed in its NEOWISE light curve between 2013 and 2021 either, no such massive dust production occurred during that time. There are possible signs of a dust-driven wind as observed in RCB stars, but this wind is not accompanied with large photometric declines, so that A166 may be in a moderate dust production phase. These on and off phases in dust production are already seen in a few known RCB stars as they are indirectly indicated by the presence of multiple circumstellar dust shells around them (see Section~\ref{sec_dust}). From its position in the HR diagram, A166 could be a newly formed HdC star that has recently started its path on the evolution track, and thus is starting its dust production phase corresponding to RCB stars.

\subsection{How many Galactic HdC stars exist?}

Based on their results of the first all-sky search for RCB stars using their mid-IR emissions, \citet{2020A&A...635A..14T} estimated that the total number of Galactic HdC stars (i.e. the sum of RCB and dLHdC stars) probably is between 300 and 500. They also discussed a long list of uncertainties that might affect these numbers. The uncertainties were mostly due to the different circumstellar environments observed around HdC stars, from dustless to very enshrouded, in some cases, even multiple shells. When no dust is observed around HdC stars, the authors used the only five dLHdC stars that were known at the time and compared this number to the number of bright RCB stars known before the modern monitoring surveys. They estimated that probably no more than 60 exist in the Galaxy. This number now needs to be revised upwards, as does the total number of Galactic HdC stars.

The RCB stars and dLHdC stars seem to share the same Galactic distribution (Section~\ref{sec_distrib}). To evaluate the number ratio between these two populations, we should consider only the sky area located towards the bulge (i.e.|\textit{l}|<45 deg and |\textit{b}|<15 deg) in the region affected by interstellar reddening E(B-V) lower than 1.0 mag, as calculated by \citet{2011ApJ...737..103S}. This corresponds to the sky area in which we completed our observation of all our dLHdC candidates. There, 53 RCB stars are known, while we found 21 dLHdC stars in addition to the 3 known nearby ones that are also located towards this region. The ratio R$_{HdC}$ in number of RCB to dLHdC stars is thus already close to 2, but we did not take the following three effects into account: 1) our search was not designed originally to find warm dLHdC stars (Section~\ref{sec_discoveries}); 2) if the intrinsic luminosity of dLHdC stars suddenly drops for a temperature below 5000 K, a selection effects towards these cool temperatures are possible; and 3) the absolute magnitude of dLHdC stars is fainter by $\sim$1.5 mag than that of RCB stars, which could imply that we did not probe the same volume with the same detection efficiency to search for the two populations.

It is already not unreasonable to imagine that the ratio R$_{HdC}$ could tend to unity simply by adjusting for the first two effects, but we recognise that this is a clear unknown. The impact of the third effect is certainly not negligible. We have shown that HdC stars have a bulge-like distribution, and we succeeded in finding dLHdC stars up to 11 kpc in the direction of the Galactic bulge. Consequently, we succeeded in efficiently probing a large fraction of the Galactic bulge towards the sky area we surveyed. The magnitude limit reached by our survey (which is constrained by our initial star list selection with G$<$15 mag and the cut on M$_G$ afterwards; see Fig.~\ref{fig_Select-GAIA}, right) is faint enough to claim that we reached a detection efficiency close to the values achieved for RCB stars surveys towards a sky area impacted by interstellar reddening lower than E(B-V)$<$0.7 mag. Above this value, our detection efficiency should decrease significantly for bulge dLHdC stars fainter than M$_V$>-4 mag. Furthermore, to select our initial GAIA eDR3 star list, we applied a strict cut on the parallax S/N (i.e. $>$3 $\sigma$). This arbitrary choice has the direct consequence to decrease our detection efficiency as a function of distance \citep{2018A&A...616A...9L}. The effect is attenuated by the brightness of the sample of stars selected (G$<$15 mag), but we should definitely expect to be less efficient in finding dLHdC stars at the distance of the bulge than in front of it (i.e. $\sim$6 kpc). We did not quantify the overall accumulated impact of the effects discussed above, but the tendency is towards a lower ratio R$_{HdC}$ value.

Overall, the initial estimate from \citet{2020A&A...635A..14T} of the total number of Galactic HdC stars should be revised upwards as they used a high value of 6 for R$_{HdC}$. This ratio should in fact reasonably range between 0.5 and 2, which would then correspond to the total number of Galactic HdC stars of between 350 and 1250 when we take the pessimistic and optimistic scenarios described by \citet{2020A&A...635A..14T} in relation to the total number of Galactic RCB stars into account. RCB stars were the tip of the iceberg in the world of HdC stars. We should now recognise that dLHdC stars might have similar numbers to RCB stars or might even outnumber them.

\section{Summary \label{sec_summary}}

After the release of the GAIA eDR3 dataset \citep{2021A&A...649A...1G}, which contains the full astrometric solution for nearly 1.5 billion sources between 3<G<21 magnitudes, we selected about 720 dLHdC star candidates using the GAIA photometry and  the near-IR photometry from the 2MASS survey in sky areas affected by interstellar reddening E(B-V) lower than 1.0 mag. We defined our pragmatic selection cuts based only on the four bright dLHdC stars known at the time to optimise our rejection rate while keeping open the possibility of finding new dLHdC stars over a wider range of temperature and brightness. We spectroscopically followed up $\sim$70\% of all our targets, the remaining ones were either unreachable with our telescope during our seven-month observation campaign, or had a very low priority as they were suspected of being blends with nearby objects. We discovered 27 new dLHdC stars and one new RCB star. We also had the fortunate surprise to reveal two new EHe stars, A208 and A798. 

We have increased the number of known dLHdC stars by a factor of 6, which allows us to better study this particular population of HdC stars and thus to characterise them in relation to their dusty counterpart, the RCB stars. First, we surprisingly  found that their hydrogen deficiency is generally less pronounced than is usually seen in the atmosphere of RCB stars, as we detected traces of weak H$_\alpha$ absorption lines in 19 of them ($\sim$63\%). This was seen in only one of the four dLHdC stars previously known, that is, HD 148839. Second, most of the new dLHdC stars show strong C$_2$ features, but only weak CN bands, indicating that their atmospheres have a lower nitrogen abundance than the RCB stars.

Third, we found that a few of the new stars are located in the Galactic halo (A166, F152, F75, C20, and HE 1015-2050), but most seem to have positions and geometric distances that support a bulge-like distribution, similar to RCB stars.

Fourth, our study of the WISE mid-IR photometry for all new HdC stars confirm that most of them are indeed dustless, and that one, C105, is a new RCB star surrounded by a typical warm circumstellar shell of $\sim$500 K. Furthermore, a few of them show some sign of a moderate dust production. This is the case for F75, F152, C526, and A166. They appear to be in a transition time, entering or leaving the RCB phase, and may therefore not be part of this dustless population of HdC stars. This is principally supported by their WISE photometry, which seems to indicate the presence of a thin layer of warm circumstellar dust shell for the first two, and a cold distant shell for the third, A166. A166 could be a supergiant HdC star that has started its path on the evolution track not long ago, and is now starting up a dust production phase like an RCB star. Evidence of a moderate amount of dust production for C526 comes from a burst lasting 5 years that was monitored by the NEOWISE mid-IR survey. Small photometric declines ($\sim$0.4 mag) were even observed in the visual light curves of F75, as well as some emission lines in its spectrum, as we did during our survey for F152.

Fifth, we studied the light curves of the new dLHdC stars and found that they either present no detectable variability at the photometric resolution of the ASAS-SN monitoring survey, or if they do, present some irregular variabilities with a weak peak-to-peak amplitude (typically $\sim$0.1 mag) and with a short timescale ($\sim$10 days). This is in clear contrast with RCB stars, which show variabilities at maximum brightness of about 0.3-0.4 mag in total amplitude and have typical timescales between 10 to 60 days.

Sixth, astonishingly, the population of dLHdC stars is clearly intrinsically fainter than the RCB star population. In a reconstructed M$_V$ versus (V-I)$_0$ colour-magnitude diagram, we found that within the colour range 0.4$<$(V-I)$_0<$1.0 mag, dLHdC stars are distributed with a median absolute magnitude M$^{dLHdC}_V\sim$-3 mag, while Galactic RCB stars have M$^{RCB}_V$ of -4.5 mag. We can even observe that the brightness of the entire population of HdC stars is distributed over 3 magnitudes for the same temperature range. This is in complete contradiction to the first view we had from simply using the Magellanic RCB stars \citep{2001ApJ...554..298A,2009A&A...501..985T}. We did not find any dLHdC stars cooler than $\sim$5000 K either, if we consider A166 to be an RCB star. We do not have a strong argument for a selection effect that could explain this missing sample of stars, and this might be a physical effect related to the evolution of this group of stars. We will focus on searching for cold dLHdC stars to test this hypothesis.

Overall, in addition to the amount of circumstellar dust surrounding them, our survey has revealed new differences between the two populations of HdC stars: hydrogen and nitrogen abundances, photometric variability, absolute magnitude, and temperature range. These need to be added to the already suspected difference, which is the lower $^{16}$O/$^{18}$O ratio observed in the atmosphere of dLHdC stars (\citet{2007ApJ...662.1220C,2009ApJ...696.1733G}; \citet{Karambelkar_2022}). The atmospheres of HdC stars, including RCB stars, are rich in the oxygen 18 isotope \citep{2007ApJ...662.1220C}, but the dustless ones appear to be the richest.

In the framework of the double-degenerate scenario, which is the favourite scenario to explain the origin of HdC stars \citep{1984ApJ...277..355W,2011MNRAS.414.3599J,2012JAVSO..40..539C}, the observed wide range in absolute magnitude (up to 3 mags) could be explained by the fact that we witness a series of evolutionary sequences of WD mergers with a wide range of initial total mass. We presented the result of a population synthesis simulation whose outcome is a plausible distribution of this initial total mass. It ranges between 0.6 and 1.05 M$_\sun$ and has a clear bimodal structure. The massive end of the mass distribution is mostly formed by the merger of a hybrid COHe WD with a CO WD. The lower the total mass, the fainter the resulting supergiant HdC star is after its evolution to warmer temperatures. These evolutionary sequences were simulated by \citet{2002MNRAS.333..121S}. They even showed that the initial phase of HdC stars, just after the merger, starts at warmer temperature for a lower total masses. This effect might explain our missing population of cold dLHdC stars. Furthermore, a noticeable difference is observed between the shape of the distribution of absolute magnitudes between the Galactic RCB stars and the Magellanic ones. They both show a similar maximum luminosity at warmer temperature but a difference up to $\sim$1.5 mag on the cooler side. Magellanic cold RCB stars are brighter. The reason might be a physical difference in the original stellar population in metallicity. \citet{2015MNRAS.450.3708R} have shown that metal-rich progenitors result in less massive WD remnants because the mass-loss rates associated with high metallicity values are higher. However, to match our observations, the \citet{2002MNRAS.333..121S} simulations show that the mass of the resulting Magellanic WDs systems needs to be similar to that of the Galactic ones. The only difference needs to be on the WDs mass ratio. This will require further investigation. We are also observing an envelope of maximum brightness versus temperature in the distribution of HdC stars in the M$_V$ versus (V-I)$_0$ colour-magnitude diagram. We interpret this as the evolutionary sequence of the maximum total WD mass possible to create an HdC star. Above this mass threshold, the WD mergers may result in supernovae \citep{2010ApJ...714L..52S,2011MNRAS.417..408R}. Finally, we note that if dLHdC stars are indeed the result of WD mergers of lower total mass than those that create RCB stars, the physical process needed to create dust from atmospheric convection is highly dependent on this initial total mass.

As a whole, our observations satisfy predictions made for the double-degenerate scenario. We have presented new and strong arguments that support this formation model for HdC stars.

We now have several pieces of evidence that show that an underlying population of real long-term dustless HdC stars exists and that they are characterised by the group of dLHdC stars of lower luminosity with M$_V\sim$-3 mag, as no RCB star is known in this luminosity and temperature range. This population of dLHdC stars is not simply a phase within the evolution of an RCB star to its subsequent stage, an EHe star, as previously thought. They probably are a different population of HdC stars that evolve on their own and that potentially produce no dust during their lifetimes, or produce dust at such a low rate that no circumstellar dust shell ever has the time to form. We will need to characterise this underlying population. Visual to mid-IR photometric monitoring could help to find those that are in a transition time into or out of the RCB phase, as we showed for F75, F152, C526, and A166. The new sample of dLHdC stars we discovered opens a new window in the quest to decipher the origin of HdC stars. RCB stars are the tip of the iceberg in that world of stars. As many dLHdC stars as RCB stars might exist in our Galaxy, if not more. Overall, we estimate the total number of Galactic HdC stars to be between 350 and 1250.

We should be able to further constrain this number with upcoming surveys that will focus on searching for the warmest and coolest dLHdC stars. Our serendipitous discovery of two EHe stars has also led to some ideas about how a dedicated search for these rare helium-rich blue supergiant stars might be searched for. Finally, we underline the importance of a future detailed abundance analysis of this population of dLHdC stars, similar to those that have been made for RCB stars \citep{2000A&A...353..287A,2012ApJ...747..102H,2017PASP..129j4202H,2021ApJ...921...52P}, to recognise possible differences between the two populations that are caused by their distinct original WD systems in age, mass, and mass ratio.
 
\begin{acknowledgements}

PT is profoundly grateful for the long unconditional support and love from his father, Claude Tisserand (1945-2020), and this paper is dedicated to his memory.\\
He personally thanks Tony Martin-Jones for his usual highly careful reading and comment. PT acknowledges also financial support from "Programme National de Physique Stellaire" (PNPS) of CNRS/INSU, France. AJR is funded through the Australian Research Council under award number FT170100243. We also thank the team located at Siding Spring Observatory that keeps the 2.3m telescope and its instruments in good shape, as well as the engineer, computer and technician teams located at Mount Stromlo Observatory that have facilitated the observations.

Palomar Gattini-IR (PGIR) is generously funded by Caltech, the Australian National University, the Mt Cuba Foundation, the Heising Simons Foundation and the Binational Science Foundation. PGIR is a collaborative project among Caltech, the Australian National University, the University of New South Wales, Columbia University and the Weizmann Institute of Science. MMK acknowledges generous support from the David and Lucille Packard Foundation. MMK and EO acknowledge the US-Israel Bi-national Science Foundation Grant 2016227. MMK and JLS acknowledge the Heising-Simons foundation for support via a Scialog fellowship of the Research Corporation. MMK and AMM acknowledge the Mt Cuba foundation. J. Soon is supported by an Australian Government Research Training Program (RTP) Scholarship.

This work has made use of data from the European Space Agency (ESA) mission
{\it Gaia} (\url{https://www.cosmos.esa.int/gaia}), processed by the {\it Gaia}
Data Processing and Analysis Consortium (DPAC,
\url{https://www.cosmos.esa.int/web/gaia/dpac/consortium}). Funding for the DPAC
has been provided by national institutions, in particular the institutions
participating in the {\it Gaia} Multilateral Agreement.

This publication makes use of data products from the Two Micron All
Sky Survey, which is a joint project of the University of Massachusetts
and the Infrared Processing and Analysis Center/California Institute of
Technology, funded by the National Aeronautics and Space Administration
and the National Science Foundation.

This publication makes use of data products from the Wide-field
Infrared Survey Explorer, which is a joint project of the University of
California, Los Angeles, and the Jet Propulsion Laboratory/California
Institute of Technology, funded by the National Aeronautics and Space
Administration.

This research has made use of the SIMBAD database,operated at CDS, Strasbourg, France.

The DASCH project at Harvard is grateful for partial support from NSF grants AST-0407380, AST-0909073, and AST-1313370.

\end{acknowledgements}

\bibliographystyle{aa}
\bibliography{HdCdiscovery}


\end{document}